\documentclass[12pt]{article}
\usepackage{amsmath,amssymb,amsfonts,amsthm,hyperref,bbm,latexsym,epsfig}
\usepackage{graphicx,multirow}
\usepackage{cite}
\usepackage{cancel}
\usepackage[width=1.1\textwidth]{caption}
\usepackage{orcidlink}

\newcommand{\bs}{\begin{subequations}}
\newcommand{\es}{\end{subequations}}
\newcommand{\be}{\begin{equation}}
\newcommand{\ee}{\end{equation}}
\newcommand{\ba}{\begin{eqnarray}}
\newcommand{\ea}{\end{eqnarray}}
\newcommand{\no}{\nonumber \\}

\newcommand{\ie}{\textit{i.e.}}
\newcommand{\viz}{\textit{viz.}}
\newcommand{\eg}{\textit{e.g.}}
\newcommand{\cf}{\textit{cf.}}
\newcommand{\vs}{\textit{vs.}}

\newcommand{\gfc}{G_{F(\mathrm{charged})}}
\newcommand{\gfchat}{\widehat G_{F(\mathrm{charged})}}

\allowdisplaybreaks

\begin{document}

\title{Oblique corrections
when $m_W \neq m_Z \cos{\theta_W}$ at tree level}

\author{
  \addtocounter{footnote}{2}
  Simonas Drauk\v{s}as,\orcidlink{0000-0003-4796-2760}$^{(1)}$\thanks{E-mail:
    {\tt simonas.drauksas@ff.vu.lt}}
  \
  Vytautas D\={u}d\.{e}nas,\orcidlink{0000-0001-9405-9959}$^{(1)}$\thanks{E-mail:
    {\tt vytautas.dudenas@tfai.vu.lt}}
  \ and\!\! \addtocounter{footnote}{1}
  Lu\'\i s Lavoura\orcidlink{0000-0002-7929-9871}$^{(2)}$\thanks{E-mail:
    {\tt balio@cftp.tecnico.ulisboa.pt}}
  \\*[3mm]
  $^{(1)} \! $
  \small Institute of Theoretical Physics and Astronomy, Faculty of Physics,
  Vilnius University, \\
  \small 9 Saul\.{e}tekio, LT-10222 Vilnius, Lithuania
  \\*[2mm]
  $^{(2)} \! $
  \small CFTP, Instituto Superior T\'ecnico, Universidade de Lisboa, \\
  \small Av.~Rovisco Pais~1, 1049-001 Lisboa, Portugal
}
\date{\today}

\maketitle

\begin{abstract}
The parametrization of the oblique corrections through $S$,
$T$,
and $U$---later extended by $V$,
$W$,
and $X$---is a convenient way of comparing the predictions
for various electroweak observables at the one-loop level
between the Standard Model and its extensions.
That parametrization assumes that the extensions under consideration
have ${SU(2)\times U(1)}$ gauge symmetry
\emph{and} the tree-level relation $m_W = m_Z \cos{\theta_W}$
between the Weinberg angle and the gauge-boson masses.
In models where that relation does not hold at the Lagrangian level,
the parameter $T$ is not ultraviolet-finite,
making the parametrization inadequate. 
We present expressions
that parametrize the difference of the various predictions of
two models with $m_W \neq m_Z \cos{\theta_W}$ in terms of oblique parameters.
The parameter $T$ does not play a role in those expressions.  
Conveniently,
they may be reached
from the ones
that were derived for models with tree-level $m_W = m_Z \cos{\theta_W}$,
by performing a simple substitution for $T$.
We also discuss the difficulties in using oblique parameters
when comparing a model with $m_W \neq m_Z \cos{\theta_W}$ to the Standard Model.
Finally,
we compute the relevant five oblique parameters $S$, $U$, $V$, $W$, and $X$
in the SM extended by
both, hypercharge $Y=0$ and $Y=1$, triplet scalars. 
\end{abstract}

\pagebreak
\tableofcontents

\section{Introduction}\label{sec:intro}

In electroweak (EW) physics an important concept
is the one of oblique corrections~\cite{kennedy1988},
which are the radiative corrections to the gauge-boson two-point functions.
Those self-energies are some of the most important radiative corrections
to the scattering amplitudes of fermions.
As opposed to ``direct'' radiative corrections
(\ie, box and vertex corrections),
oblique corrections are process-independent,
\ie\ they contribute in the same way to various amplitudes
without the need to recalculate them for every single process.
Precision measurements in the EW sector
can probe the oblique corrections for hints of physics
beyond the Standard Model (SM),
\ie\ New Physics (NP).

In~\cite{veltman1977} it was noticed that
at low energies the effects of particles with masses much larger
than those of the gauge bosons $W^\pm$ and $Z^0$ may be visible.
In particular,
the ratio between the charged and the neutral weak currents---the Veltman
$\rho$ parameter---is sensitive to such non-decoupling effects.
The non-decoupling effects in the oblique corrections
were discussed more systematically
in~\cite{peskin1990,kennedy1990,holdom1990,golden1991},
and quantities other than the $\rho$ parameter were shown to be affected too.
These heavy-physics effects are most commonly captured by the $S$,
$T$,
and $U$ oblique parameters (OPs) introduced in~\cite{peskin1990,peskin1992};
they are used to parametrize the oblique corrections
to various observables in the EW sector.
Equivalent formulations are available
in~\cite{altarelli1991,altarelli1992,marciano1990,kennedy1991}.

The $STU$ parameters are not appropriate
to account for light-physics effects in the oblique corrections;
hence,
an extension via additional parameters $V$,
$W$,
and $X$ was introduced in~\cite{maksymyk1994,burgess1994}.
The $STUVWX$ formalism covers all the EW observables
and is valid irrespective of whether NP
is heavier or lighter than the Fermi scale.
This formalism seems to be the most versatile one
for parametrizing the oblique corrections,
and we focus on it in this paper. 

We note that another generalization of the $STU$ parameters exists,
which introduces four additional OPs
(confusingly called $XWVY$) in~\cite{barbieri2004}. 
The parametrization of~\cite{barbieri2004},
however,
is less general,
as it assumes that NP is much heavier than the EW scale. 
We relate the $STUVWX$ parameters used in this paper
with the ones of~\cite{barbieri2004},
in the limit of only heavy NP,
in Appendix~\ref{app: barbieri}. 
In that limit,
the parameters of~\cite{barbieri2004} may also be related
to the Wilson coefficients of
the SM Effective Field Theory (SMEFT)~\cite{barbieri2004,wells2016}.

It is worth mentioning that nowadays the EFT approach
is often utilized and may be related to the parametrizations
of the oblique corrections,
see \textit{e.g.}~\cite{barbieri2004,fan2022,bagnaschi2022}.
An unambiguous relation to the OPs is possible for restricted EFTs,
\viz\ when one assumes that the heavy new physics
contributes to the observables only via oblique corrections
(the so-called ``universal theories''~\cite{wells2016}). 
EFT does offer a very general and systematic approach
to the parametrization of NP,
also beyond the EW sector;
in particular,
the ``direct'' radiative corrections are not neglected in EFT.
The downside of EFT lies precisely in its generality:
the number of parameters that one needs to fit to the data
is huge---for example,
in the most general case there are 2499 dimension-six
operators
in SMEFT~\cite{alonso2014,henning2017}.
For this reason,
the OPs are handy,
and indeed they continue to be widely used in testing models of NP.
Moreover,
if NP is not far away from the EW scale,
then the EFT approximation becomes less precise
and the OPs may give a more accurate description~\cite{banerjee2023}. 

The six OPs $S, T, \ldots, X$
consider the difference between two models:
the SM and an extension thereof with extra matter
(\ie, fermions and/or scalars)
that we call the New Physics Model (NPM).\footnote{Models of NP
with extra gauge bosons in the EW sector
must be renormalized differently from the SM and,
except in some specific cases~\cite{barbieri2004},
can \emph{not} in general be treated through OPs.}
Both the SM and the NPM have $SU(2)\times U(1)$ gauge symmetry.
Moreover,
they both have a so-called ``custodial symmetry'' that relates $m_W$ and $m_Z$
(the masses of the gauge bosons $W^\pm$ and $Z^0$,
respectively)
at the Lagrangian level. 
This is often expressed in terms
of the Veltman $\rho$ parameter~\cite{veltman1977} as
\be
\widehat \rho \equiv \frac{ \widehat m_W^2}{\widehat c^2 \widehat m_Z^2} 
\begin{cases} 
= 1 & \text{with custodial symmetry}, \\
\text{is not fixed} &\text{without custodial symmetry},
\end{cases}
\label{eq:Veltman rho hat}
\ee
where $\widehat c \equiv \cos{\widehat \theta_W}$
is the cosine of the Weinberg angle $\widehat \theta_W$.
In Eq.~\eqref{eq:Veltman rho hat} and hereinafter,
we use hats to denote tree-level (bare) parameters,
and we drop the hats in (one-)loop-level parameters.
Equation~\eqref{eq:Veltman rho hat} gets loop corrections
and the value of $\rho$ at loop level depends
on the renormalization scheme---or,
equivalently,
on the choice of input observables.

In models with custodial symmetry,
only three observables are needed to describe the EW sector;
we take those observables to be the fine-structure constant $\alpha$,
the Fermi coupling constant $\gfc$ that is measured in muon decay,
and $m_Z$.
In light of the new measurement of $m_W$
by the CDF Collaboration~\cite{cdfcollaboration+++2022},
which disagrees with the SM prediction by seven standard deviations,
it is important to consider EW models without custodial symmetry,
\ie\ with $\widehat \rho$ unfixed.
Then four,
instead of just three,
input observables are needed to renormalize the EW model.
In this paper we take those four observables to be the former three plus $m_W$.
Since there is one more input observable in the renormalization procedure,
one of the six parameters $S, T, \ldots, X$,
is not calculable any more from the bare self-energies only,
\ie\ it becomes renormalization scheme-dependent.
Indeed,
when $\widehat \rho$ is not fixed to $1$
the parameter $T$ is ultraviolet (UV)-divergent,
as has been affirmed before us
by multiple authors~\cite{lynn1992,blank1998,czakon2000,forshaw2001,forshaw2003,chen2006a,chankowski2007,chen2008,albergaria2022,rizzo2022}
(see also Appendix~\ref{app:divergences}).
So,
in that case there are only five oblique parameters,
\viz\ $S$,
$U$,
$V$,
$W$,
and $X$.
We emphasize that these implications
of the non-existence of custodial symmetry
were noticed already long ago,
for instance in~\cite{kennedy1989,lynn1992,peskin1992}.

The advantage of the OPs is that they render easy to compare
the predictions of two custodially symmetric models,
\viz\ the NPM and the SM.
However,
if some other base model (BM) is compared with a model beyond it (the BBM),
and neither the BM nor the BBM have $\widehat \rho$ fixed,
then the expressions that were derived
to compare the SM with an NPM no longer hold,
since they all contain $T$ which is now UV-divergent.
In order to clearly separate these cases,
throughout this paper we will use the terms SM,
NPM,
BM,
and BBM according to the following rules:
\begin{itemize}
\item The SM and the NPM have $\widehat \rho = 1$;
  the NPM is identical to the SM plus some extra matter.
\item The BM and the BBM do not have fixed $\widehat \rho$;
  the BBM is identical to the BM plus some extra matter.
\end{itemize}
For comparing the EW predictions
between two models of interest there are three cases
that could be relevant and that need to be studied separately:
\begin{enumerate}
\item SM \vs\ NPM,
  \textit{i.e.}\ two models that have $\widehat{\rho}=1$.
\item BM \vs\ BBM,
  \textit{i.e.}\ two models that have free $\widehat{\rho}$.
\item SM \vs\ BBM,
  \textit{i.e.}\ one model has $\widehat{\rho}=1$
  and the other one has $\widehat{\rho}$ free.
\end{enumerate}
The first case was studied in~\cite{peskin1992},
where the OPs were initially introduced. 
To the best of our knowledge,
a genuine formalism of OPs
has not yet been developed for the two remaining cases. 
The third case has seen attempts at it---for example,
there are approaches wherein the parameter $T$ is modified
to include tree-level contributions,
while the loop contributions are computed
in the limit without custodial symmetry breaking---see
\textit{e.g.}~\cite{forshaw2001,chankowski2007,cheng2023,song2023}.
Of course,
this implies that $T$ becomes renormalization scheme-dependent
and should be renormalized,
which is not the case in the original theory of OPs.
For example,
such an approach was taken in~\cite{burdman2008},
where the parameter $S$ is UV-divergent.

It is most important to highlight
that the applicability of OPs to EW models without custodial symmetry
is limited.
A consistent treatment is sometimes not employed;
indeed,
it is not uncommon to find the usual OPs,
wherein custodial symmetry is assumed,
being applied to models without that symmetry~\cite{chun2012,mandal2022,
  hagiwara1994,heeck2022,cheng2023,song2023}. 

In this paper we produce a formalism of OPs for the second case,
\viz\ a BBM \vs\ the BM.
We also outline the issues that one faces
when trying to deal with the third case,
which we leave as an open problem.
The third case would be the most useful generalization
of the formalism,
because the current base model is the SM that has $\widehat \rho =1$. 
However,
the comparison of BM \vs\ BBM is also significant:
if one calculates the full one-loop EW predictions of some BM,
then the possibility to relate them to the predictions of any BBM
through OPs allows skipping the full calculation of those predictions
in the BBM.
Moreover,
it is not unthinkable that,
in the future,
the standard model of particle physics will have $\widehat \rho$ free,
and then the second case would become immediately relevant.

\subsection{Main definitions in the
  \texorpdfstring{$SU(2) \times U(1)$}{SU(2)xU(1)} gauge theory}

All EW models considered in this paper
have $SU(2) \times U(1)$ gauge symmetry. 
We shortly review the main tree-level conventions and definitions
for those models.

The covariant derivative is written
\be
D_\mu = \partial_\mu - i\, \widehat g\, \sum_{a=1}^3 T^a W_\mu^a
- i\, \widehat g^{\, \prime}\, Y_H B_\mu,
\ee
where the gauge coupling constants $\widehat g$ and $\widehat g ^{\, \prime}$
of $SU(2)$ and $U(1)$,
respectively,
are positive,
the $T^a$ are the generators of $SU(2)$,
and $Y_H$ is the (weak) hypercharge,
\ie\ the generator of $U(1)$.
For the sake of brevity,
we omit the hats over the gauge fields,
\viz\  we write {$B_\mu$} instead of {$\widehat B_\mu$}, and so on.
For matter
(\ie, either fermions or scalars)
placed in doublets of $SU(2)$,
$T^a = \sigma^a / 2$,
where the $\sigma^a$ are the Pauli matrices.

After EW symmetry breaking,
the photon field $A_\mu$ and the $Z^0$-boson field $Z_\mu$
are related to the gauge-eigenstate fields $W^3_\mu$ and $B_\mu$ by
\be
\begin{array}{rcl}
  A_\mu &=& \widehat c\, B_\mu + \widehat s\, W^3_\mu
  \\*[1mm]
  Z_\mu &=&\widehat  c\, W^3_\mu - \widehat s\, B_\mu
\end{array}
\quad \Leftrightarrow \quad
\begin{array}{rcl}
  B_\mu &=& \widehat c\, A_\mu - \widehat s\, Z_\mu
  \\*[1mm]
  W^3_\mu &=& \widehat c\, Z_\mu + \widehat s\, A_\mu
\end{array},
\label{eq:Weinberg rotation}
\ee
where $\widehat s$ and $\widehat c$ are the sine and the cosine,
respectively,
of the Weinberg angle $\widehat \theta_W$.
The $W^\pm$-boson field is
\be
W_\mu^\pm = \frac{W_\mu^1 \mp i W_\mu^2}{\sqrt{2}}.
\ee
The absolute value of the electric charge $\widehat e$
is related to $\widehat g$,
$\widehat g^\prime$,
$\widehat \theta_W$,
and the fine-structure constant $\widehat \alpha$ by
\be
\widehat e = \widehat g\, \widehat s
= \widehat g^{\, \prime}\, \widehat c
= \sqrt{4 \pi  \widehat \alpha}.
\ee

In the SM and in an NPM only scalar fields that have weak isospin $1/2$,
\ie\ doublets of $SU(2)$,
and have weak hypercharge $\pm 1/2$ get vacuum expectation values
(VEVs).\footnote{Possibly,
scalar fields with null isospin and hypercharge acquire VEVs too,
but those VEVs neither break {$SU(2) \times U(1)$}
nor contribute to {$m_W$} and {$m_Z$}.}
As a consequence,
at tree level the masses of the gauge bosons are related through
\be
\widehat m_W = \widehat c\, \widehat m_Z.
\label{eq:Veltman rho hat=1}
\ee
This is due to the accidental `custodial' $SU(2)$ symmetry
of the scalar potential~\cite{lynn1992}
when all the scalars with VEVs are in $SU(2$) doublets.
At loop level Eq.~\eqref{eq:Veltman rho hat=1} gets corrections,
because
the custodial symmetry is not a symmetry of the complete Lagrangian:
both the hypercharge gauge interactions
and the Yukawa couplings break it.

An important equation relates the tree-level
$\widehat G_{F(\mathrm{charged})}$ to the Weinberg angle:
\be
\widehat s^{\, 2} = \frac{\pi \widehat \alpha}{\sqrt{2}\,
  \widehat G_{F(\mathrm{charged})}\, \widehat m_W^2}\, .
\label{eq:s2 hat} 
\ee
When there is custodial symmetry,
\viz\ when Eq.~\eqref{eq:Veltman rho hat=1} holds,
one may use instead
\be
\widehat s^{\, 2}\, \widehat c^{\, 2}
= \frac{\pi \widehat \alpha}{\sqrt{2}\,
  \widehat G_{F(\mathrm{charged})}\, \widehat m_Z^2} \,.
\label{eq:s2c2 hat}
\ee
Any of the three Eqs.~\eqref{eq:Veltman rho hat=1},
\eqref{eq:s2 hat},
and~\eqref{eq:s2c2 hat}
may be used as the \emph{definition} of $\widehat \theta_W$
in models with $\widehat\rho=1$; 
which definition one uses partially depends
on which input observables one chooses.
However,
Eqs.~\eqref{eq:Veltman rho hat=1} and~\eqref{eq:s2c2 hat}
do not hold when $\widehat \rho \neq 1$;
in that case,
only Eq.~\eqref{eq:s2 hat} may be used to define $\widehat \theta_W$.

\subsection{Definitions of the oblique parameters}

The oblique corrections
originate in the bare self-energies of the gauge bosons.
Those self-energies are decomposed as
\be
\Pi^{\mu\nu}_{V V^\prime} \left( p \right) =
g^{\mu\nu}\, \Pi_{V V^\prime} \left( p^2 \right) + p^\mu p^\nu\, \text{terms},
\ee
where $V$ and $V^\prime$ are the vector bosons,
$\mu$ and $\nu$ are their polarizations,
and $p^\mu$ is their four-momentum.
The OPs do not depend on the coefficients of the $p^\mu p^\nu$ terms;
all the OPs are expressed in terms
of the functions $\Pi_{V V^\prime} \left( p^2 \right)$.
Starting from those functions,
we define
\bs
\label{eq:NP}
\ba
\widetilde \Pi_{V V^\prime} \left( p^2 \right)
&\equiv&
\frac{\Pi_{V V^\prime} \left( p^2 \right)
  - \Pi_{V V^\prime} \left( 0 \right)}{p^2}, \label{eq:self tilde}
\\
\Pi_{V V^\prime}^\prime \left( p^2 \right)
&\equiv&
\frac{\mathrm{d} \Pi_{V V^\prime} \left( p^2 \right)}{\mathrm{d} p^2}.
\label{eq:self prime}
\ea
\es
The functions $\Pi_{V V^\prime} \left( p^2 \right)$,
$\widetilde \Pi_{V V^\prime} \left( p^2 \right)$,
and $\Pi_{V V^\prime}^\prime \left( p^2 \right)$
depend on the matter content of each model.
We place a superscript `M' on each of them
to identify the model in which they are computed.

We define the `model parameters' $S^\mathrm{M}, \ldots, X^\mathrm{M}$
as in~\cite{maksymyk1994}:\footnote{
We include $\Pi^\mathrm{M}_{ZA} \left( 0 \right)$
in the definition of $T^\mathrm{M}$.
This quantity vanishes if only the BSM contribution to it is considered,
so it is usually omitted from the definition of $T$,
for instance in~\cite{maksymyk1994}.}
\bs
\label{eq:STU definitions}
\ba
S^\mathrm{M} &=&
\frac{4 s^2 c^2}{\alpha}
\left[ \widetilde{\Pi}^\mathrm{M}_{ZZ} \left( m_Z^2 \right) 
+ \frac{s^2 - c^2}{sc}\, \Pi_{ZA}^{\prime \, \mathrm{M}} \left( 0 \right) 
- \Pi_{AA}^{\prime\, \mathrm{M}} \left( 0 \right)
\right],
\label{eq:S in mass}
\\
T^\mathrm{M} &=&
\frac{1}{\alpha} \left[
  \frac{\Pi^\mathrm{M}_{WW}\left( 0 \right)}{m_W^2}
  - \frac{\Pi_{ZZ}^\mathrm{M}\left( 0 \right)}{m_Z^2}
  - \frac{2 s}{c}\, \frac{\Pi^\mathrm{M}_{ZA} \left( 0 \right)}{m_Z^2}\right],
\label{eq:T in mass}
\\
U^\mathrm{M} &=& 
\frac{4 s^2}{\alpha} \left[
  \widetilde{\Pi}^\mathrm{M}_{WW} \left( m_W^2 \right)
  - c^2\, \widetilde{\Pi}^\mathrm{M}_{ZZ} \left( m_Z^2 \right)
  - 2 s c\, \Pi_{ZA}^{\prime\, \mathrm{M}} \left( 0 \right)
- s^2\, \Pi_{AA}^{\prime\, \mathrm{M}} \left( 0 \right) \right],
\label{eq:U in mass}
\\
V^\mathrm{M} &=& \frac{1}{\alpha} \left[
  \Pi_{ZZ}^{\prime\, \mathrm{M}} \left( m_Z^2 \right)
  - \widetilde{\Pi}^\mathrm{M}_{ZZ}\left( m_Z^2 \right) \right],
\label{V in mass}
\\
W^\mathrm{M} &=& \frac{1}{\alpha} \left[
  \Pi_{WW}^{\prime\, \mathrm{M}} \left( m_W^2 \right)
  - \widetilde{\Pi}^\mathrm{M}_{WW} \left( m_W^2 \right) \right],
\label{W in mass}
\\
X^\mathrm{M} &=&
\frac{s c}{\alpha} \left[
  \Pi_{ZA}^{\prime\, \mathrm{M}} \left( 0 \right)
  - \widetilde{\Pi}^\mathrm{M}_{ZA} \left( m_Z^2 \right) \right],
\label{eq:X in mass} \\
Y^\mathrm{M} &=&
\frac{1}{\alpha} \left[
  \Pi_{AA}^{\prime\, \mathrm{M}} \left( 0 \right)
  - \widetilde{\Pi}^\mathrm{M}_{AA} \left( m_Z^2 \right) \right].
\label{eq:Y in mass}
\ea
\es
For completeness we have also introduced the parameter $Y^\mathrm{M}$,
which in~\cite{maksymyk1994} has been assumed to be small.
That parameter is used in Appendix~\ref{app: barbieri},
where it allows for a relation with a different set of parameters
defined in~\cite{barbieri2004}.
We follow~\cite{maksymyk1994} and neglect $Y^\mathrm{M}$ everywhere else.

We introduce an abbreviation for the often-appearing linear combination
\bs
\label{eq:Kustodial par}
\ba
K^\mathrm{M} &\equiv& \frac{S^\mathrm{M}}{2 c^2}
+ \frac{s^2 - c^2}{4 s^2 c^2}\, U^\mathrm{M}
\\  &=& \frac{1}{\alpha} \left[
  \frac{s^2 - c^2}{c^2}\, \widetilde{\Pi}^\mathrm{M}_{WW} \left( m_W^2 \right)
  + \widetilde{\Pi}^\mathrm{M}_{ZZ} \left( m_Z^2 \right)
  - \frac{s^2}{c^2}\, \Pi_{AA}^{\prime\, \mathrm{M}} \left( 0 \right) \right].
\ea
\es

The oblique parameter $O = S, \ldots, Y$
is most commonly defined
as the \emph{subtraction} between the model parameters
$O^{\mathrm{M}^\prime}$ and $O^\mathrm{M}$:
\begin{equation}
  \label{mvoift00}
  O \equiv O^{\mathrm{M}^\prime} - O^\mathrm{M},
\end{equation}
where the model M$^\prime$ is an extension of the model M,
\ie\ $M^\prime$ is M plus some extra scalar and/or fermion fields.
We emphasize that the model parameters in Eqs.~\eqref{eq:STU definitions}
contain contributions from loops of gauge bosons,
Goldstone bosons,
and ghosts;
therefore,
in general they are neither gauge-independent nor finite.
Contrary to the model parameters,
the OPs are both gauge-independent and finite.

\subsection{Plan of the paper}

In Section~\ref{sec2} we derive the prescription
for substituting the parameter $T$ when comparing a BBM with a BM.
In Section~\ref{sec:example} we give formulas for all the OPs,
except $T$,
when the BBM is the SM plus \emph{two} triplets of scalars,
one of them with $Y_H=0$ and the other one with $Y_H=1$,
that both have VEVs in their neutral components.
Section~\ref{sec4} discusses the difficulties in using OPs
for comparing the SM with a model where $\widehat \rho$ is free.
Section~\ref{sec:conclusions} summarizes the conclusions of our research.
Afterwards there is a set of appendices:
Appendix~\ref{app:observable list} lists the observables
that one compares between M$^\prime$ and M;
Appendix~\ref{app:NPM vs SM} derives the formulas for the comparison
of the observables between the SM and an NPM;
Appendix~\ref{app:BBM vs BM} does the same job
for the comparison between a BM and a BBM;
Appendix~\ref{app: barbieri} gives relations between the
oblique parameters of~\cite{maksymyk1994} and~\cite{barbieri2004};
Appendix~\ref{app:divergences} discusses the finiteness of the OPs;
Appendix~\ref{app:PaVe} defines the Passarino--Veltman functions
utilized in Section~\ref{sec:example}.

\section{Observables in terms of the oblique parameters}\label{sec2}

\subsection{Observables \vs\ bare parameters}

Observables are calculated from bare parameters,
at a given order in perturbation theory,
by computing the loop corrections to them. 
One then expresses the bare parameters
in terms of some chosen input observables,
and one plugs them into the expressions
for the observables that one wants to predict. 
In that way,
the procedure of defining counterterms is circumvented,
because everything is expressed in terms of observable quantities
and loop functions that are renormalization scheme-independent,
as was originally proposed in~\cite{kennedy1989}. 
The final expressions of course depend on the choice of input observables.

One of the input observables is the fine-structure constant $\alpha$,
which at one-loop level is given by
(see \eg~\cite{denner1993,pokorski2000,bohm2001,Denner2020})
\be
\alpha = \widehat \alpha^\mathrm{M} \left[
  1 + \Pi^{\mathrm{M}\prime}_{AA} \left( 0 \right)
  + \frac{2 s}{c}\, \frac{\Pi^\mathrm{M}_{ZA} \left( 0 \right)}{m_Z^2}
  \right]. \label{eq:alpha renorm}
\ee
If one considers a model M$^\prime$ with extra fermions or extra scalars,
then one always has
\begin{equation}
  \Pi_{ZA} \left( 0 \right) \equiv \Pi_{ZA}^\mathrm{M^\prime} \left( 0 \right) -
  \Pi_{ZA}^\mathrm{M} \left( 0 \right)=0.
  \label{za0}
\end{equation}
This is why $\Pi_{ZA} \left( 0 \right)$ is often omitted
when only the contributions of NP are considered. 
But,
both $\Pi_{ZA}^\mathrm{M} \left( 0 \right)$
and $\Pi_{ZA}^\mathrm{M^\prime} \left( 0 \right)$ are nonzero,
which leads to a slight complication, 
since they appear in the propagators of the $Z_\mu$ and $A_\mu$ gauge fields
and thus contribute to the $2\times 2$ mass matrix of the $Z^0$--photon system
at $p^2=0$.
To make sure that the physical photon is massless,
this term must cancel out in physical amplitudes
which include the virtual photon.
One way to deal with it is to simply renormalize everything
with an on-shell (OS) counterterm,
as in~\cite{altarelli1989}. 
Since we rather want to express everything in terms of bare self-energies,
we instead follow Kennedy and Lynn~\cite{kennedy1988},
who proposed to redefine the bare $SU(2)$ coupling
to account for the universal vertex corrections. 
Thus,
we shift the $SU(2)$ coupling through
\be
\widehat{g}^\mathrm{M} \to \widehat{g}^\mathrm{M}  \left( 1
- \frac{1}{s c}\, \frac{\Pi^\mathrm{M}_{ZA} \left( 0 \right)}{m_Z^2} \right).  \label{eq:KL redefinition}
\ee
This leads to the following redefinitions of the self-energies:
\bs
\label{eq:KL all redefinitions}
\ba
\Pi^\mathrm{M}_{ZA} \left( p^2 \right) &\to&
\Pi^\mathrm{M}_{ZA} \left( p^2 \right) -\Pi^\mathrm{M}_{ZA} \left( 0 \right),  \\
\Pi^\mathrm{M}_{ZZ} \left( p^2 \right) &\to&
\Pi^\mathrm{M}_{ZZ}\left( p^2 \right) - \frac{2 c}{s}\,
\Pi^\mathrm{M}_{ZA} \left( 0 \right), \\
\Pi^\mathrm{M}_{WW} \left( p^2 \right) &\to&
\Pi^\mathrm{M}_{WW} \left( p^2 \right) - \frac{2 c}{s}
\left( \frac{m_W^2}{c^2 m_Z^2} \right) \Pi^\mathrm{M}_{ZA} \left( 0 \right),
\ea
\es
and cancels off the $\Pi_{ZA}$ term in Eq.~\eqref{eq:alpha renorm}.
Including this redefinition,
we now express the one-loop corrections to the four observables as 
\bs
\label{eq:all inputs}
\ba
m_W^2 &=& \left(\widehat m_W^\mathrm{M}\right)^2
\left[ 1 + \frac{\Pi^\mathrm{M}_{WW} \left( m_W^2 \right) }{m_W^2}
  - \frac{2}{sc}\, \frac{\Pi^\mathrm{M}_{ZA} \left( 0 \right)}{m_Z^2}  \right],
\label{eq:mw_input}
\\
m_Z^2 &=& \left(\widehat m_Z^\mathrm{M}\right)^2
\left[ 1 + \frac{\Pi^\mathrm{M}_{ZZ} \left( m_Z^2 \right)}{m_Z^2}
  - \frac{2 c}{s}\, \frac{ \Pi^\mathrm{M}_{ZA} \left( 0 \right)}{m_Z^2} \right],
\label{eq:mz input} 
\\
\label{gfc}
G_{F(\mathrm{charged})} & = & \widehat G_{F(\mathrm{charged})}^\mathrm{M}  
\left[ 1 - \frac{\Pi^\mathrm{M}_{WW} \left( 0 \right)}{m_W^2}
  + \frac{2}{sc}\, \frac{\Pi^\mathrm{M}_{ZA} \left( 0 \right)}{\, m_Z^2}
  + \delta^\mathrm{M}_{Gc} \right], \hspace{20pt} 
\label{eq:GF charged input}
\\
\alpha &=& \widehat \alpha^\mathrm{M}
\left[ 1 + \Pi^{\mathrm{M}\prime}_{AA} \left( 0 \right) \right].
\label{eq:e_input}
\ea
\es
Note that $G_{F(\mathrm{charged})}$ is extracted from the decay rate of the muon;
$\delta^\mathrm{M}_{Gc}$ stands for the one-loop vertex,
fermion line,
and box corrections to that decay.

Because $\widehat \rho =1$ in the SM and in an NPM,
in those models just three observables,
out of the four in Eqs.~\eqref{eq:all inputs},
are needed as input.
We choose the input observables for the
\begin{equation}\label{eq: SM_NPM_inputs}
    \text{SM and an NPM:}\quad m_Z,\
  G_{F(\mathrm{charged})},\ \text{and}\ \alpha.
\end{equation}
The value of $m_W$ can then be \emph{predicted}
by using Eqs.~\eqref{eq:Veltman rho hat=1} and~\eqref{eq:s2c2 hat}.
Alternatively,
one may say that it is $\rho$ that is predicted.
Since $m_W$ is a prediction of each model,
we will write $\left( m_W^2 \right)^\mathrm{M}$
instead of just $m_W^2$ in the left-hand side of Eq.~\eqref{eq:mw_input}.

Since a BM and a BBM do not have fixed $\widehat \rho$,
they have additional freedom;
then,
all four observables in Eq.~\eqref{eq:all inputs} are needed as input,
\ie\ we utilize the input observables for a
\begin{equation}\label{eq: BM_BBM_inputs}
  \text{BM and a BBM:} \quad m_W,\  m_Z,\
  G_{F(\mathrm{charged})},\ \text{and}\ \alpha.
\end{equation}
In this case $m_W$ can \emph{not} be predicted;
but other EW observables,
that are related to these four by tree-level relations,
can.

To describe the EW observables at LEP energies,
one must take into account the running of $\alpha$~\cite{altarelli1989}.
After applying the redefinitions of Eq.~\eqref{eq:KL all redefinitions},
the fine-structure constant at the $Z^0$ pole,
\viz\ $\alpha \left( m_Z^2 \right)$,
is related to the corresponding bare parameter by
\begin{equation}
  \alpha^{\mathrm{M}} \left( m_Z^2 \right)
  = \widehat \alpha^\mathrm{M} \left[
    1 + \frac{\Pi_{AA}^\mathrm{M}\left(m_Z^2\right)}{m_Z^2} \right].
  \label{eq:alpha m_Z renorm}
\end{equation}
We add the label M to indicate that the extraction
of $\alpha \left( m_Z^2 \right)$ is model-dependent. 
Since $\alpha \left( 0 \right)$
is an input parameter, it is the same for all models; therefore, expressing
the bare parameter $\widehat \alpha^\mathrm{M}$ from Eq.~\eqref{eq:e_input},
plugging it into Eq.~\eqref{eq:alpha m_Z renorm}
both for $\mathrm{M}=\mathrm{NPM}$ and $\mathrm{M}=\mathrm{SM}$,
and taking the ratio of the two yields
\begin{equation}
  \frac{\alpha^\mathrm{NPM} \left( m_Z^2 \right)}{\alpha^\mathrm{SM}
    \left( m_Z^2 \right)} = 1 - \alpha Y.
\end{equation}
In practice,
as input one uses the SM-predicted $\alpha^\mathrm{SM} \left( m^2_Z \right)$,
\ie\ one assumes that $\alpha \left( m^2_Z \right)$ in the NPM
is approximately equal to $\alpha^\mathrm{SM} \left( m^2_Z \right)$.
This is equivalent to assuming that $Y$ is small,
as was done in~\cite{maksymyk1994}.
We shall follow this practice
and we will always consider $\alpha^\mathrm{SM} \left( m^2_Z \right)$
to be the input in our expressions for all the models
(SM, NPM, BM, and BBM). 

Since many expressions depend on the Weinberg angle,
it is convenient to define that angle in terms of the input observables. 
If one considers a BM and a BBM with Eq.~\eqref{eq: BM_BBM_inputs},
then the Weinberg angle $\theta_W$ is \emph{defined} from the input as
\begin{equation}\label{eq:s2 nohat}
  s^2  \equiv \frac{\pi\, \alpha(m_Z^2)}{\sqrt{2}\,
    G_{F(\mathrm{charged})}\, m_W^2},
\end{equation}
where $\alpha \left( m_Z^2 \right) \equiv
\alpha^\mathrm{SM} \left( m^2_Z \right)$ like we said before.
Equation~\eqref{eq:s2 nohat} is equivalent to promoting
the tree-level expression in Eq.~\eqref{eq:s2 hat}
and making it hold at loop level.
In the case with $\widehat \rho=1$
and the input observables in Eq.~\eqref{eq: SM_NPM_inputs},
which do \emph{not} include $m_W$,
one \emph{defines} the Weinberg angle $\bar \theta_W$ through
\begin{equation}
  \label{eq:s2c2 nohat}
  \bar c^2 \bar s^2  \equiv
  \frac{\pi\, \alpha \left( m_Z^2 \right)}{\sqrt{2}\,
    G_{F(\mathrm{charged})}\, m_Z^2},
\end{equation}
where $\bar c \equiv \cos{\bar \theta_W}$
and $\bar s \equiv \sin{\bar \theta_W}$.
Note that $\bar \theta_W$ defined by Eq.~\eqref{eq:s2c2 nohat}
is different from $\theta_W$ defined by Eq.~\eqref{eq:s2 nohat}
already at the one-loop level in models with $\widehat \rho=1$.
For models with free $\widehat \rho$,
the definition in Eq.~\eqref{eq:s2c2 nohat} is just not correct.

A third definition is possible when $\widehat \rho=1$,
\viz\ one may define the Weinberg angle $\widetilde \theta_W$ through
\be
\label{jb00fr}
\widetilde c^{\, 2} = \frac{m_W^2}{m_Z^2},
\ee
where $\widetilde c \equiv \cos{\widetilde \theta_W}$.
This definition is used in the OS renormalization scheme---for instance
in~\cite{Aoki1982,denner1993}---where $m_W$,
$m_Z$,
and $\alpha$ are used as input observables.
However,
as far as we know,
this definition is not often employed in the context of OPs
and therefore we consider here only the two definitions
in Eqs.~\eqref{eq:s2 nohat} and~\eqref{eq:s2c2 nohat}
but not the one in Eq.~\eqref{jb00fr}.

\subsection{The Veltman \texorpdfstring{$\rho$}{rho} parameter}

The definition of the Veltman $\rho$ parameter,
just as the definition of the Weinberg angle,
depends on the input observables that one uses. 
Thus,
using either Eq.~\eqref{eq:s2 nohat} or Eq.~\eqref{eq:s2c2 nohat}
one obtains two different definitions:
\bs
\label{eq:rho definitions}
\ba
\label{eq:s2c2 definition}
\bar\rho^\mathrm{M} &\equiv&
\frac{\left( m_W^2 \right)^\mathrm{M}}{\bar c^2 m_Z^2},
\\
\label{eq:s2 definition}
\rho &\equiv& \frac{m_W^2}{c^2 m_Z^2}.
\ea
\es
The definition in Eq.~\eqref{eq:s2c2 definition}
is only possible when $\widehat \rho = 1$,
while the definition in Eq.~\eqref{eq:s2 definition}
is possible for any $\widehat \rho$.

If one has a model M with $\widehat \rho =1$,
then one may plug Eqs.~\eqref{eq:all inputs}
into the definition of $\bar \rho$ in Eq.~\eqref{eq:s2c2 definition}.
One then groups the self-energies into model parameters
by using Eq.~\eqref{eq:STU definitions} and Eq.~\eqref{eq:Kustodial par}.
The procedure is explicitly carried out in Appendix~\ref{app:NPM vs SM},
\cf\ Eq.~\eqref{kb0007}.
One gets
\be
\label{eq:rhobar SM}
\bar \rho^\mathrm{M} = \widehat \rho^{\,\mathrm{M}}\,
\left[ 1 + \frac{\bar c^2}{\bar c^2 - \bar s^2}
\left( \alpha T^\mathrm{M} - \alpha K^\mathrm{M} \right)
+ \frac{\bar s^2}{\bar s^2 - \bar c^2}\, \delta_{Gc}^\mathrm{M}
\right],
\ee
where $\delta_{Gc}^\mathrm{M}$ is the value of $\delta_{Gc}$
computed in the model M
and $\widehat \rho^\mathrm{M} = 1$ should be implicitly understood.
Dividing Eq.~\eqref{eq:rhobar SM} for an NPM
by the same equation for the SM,
we obtain
\be
\frac{\bar{\rho}^{\, \mathrm{NPM}}}{\bar{\rho}^{\,\mathrm{SM}}}
= \frac{\widehat \rho^{\, \mathrm{NPM}}}{\widehat \rho^{\, \mathrm{SM}}}
\left[ 1 + \frac{\bar{c}^2}{\bar{c}^2 - \bar{s}^2}
  \left( \alpha T- \alpha K \right)
  + \frac{\bar{s}^2}{\bar{s}^2 - \bar{c}^2}
  \left( \delta_{Gc}^\mathrm{NPM} - \delta_{Gc}^\mathrm{SM} \right) \right].
\label{eq:rhobar NPM-SM}
\ee
We now \emph{assume} that~\cite{peskin1992}
\be
\delta_{Gc}^\mathrm{M^\prime} - \delta_{Gc}^\mathrm{M}
\ \mathrm{is\ negligible}.
\label{za1}
\ee
This is the crucial assumption in the philosophy of OPs.
We thus keep only the \emph{oblique} corrections
in Eq.~\eqref{eq:rhobar NPM-SM},
\viz
\begin{equation}
  \bar{\rho}^{\, \mathrm{NPM}} = \bar{\rho}^{\, \mathrm{SM}}
  \left[ 1 + \frac{\bar c^2}{\bar c^2 - \bar s^2}
    \left( \alpha T - \alpha K \right) \right],
  \label{eq:rhobar SM-NPM oblique}
\end{equation}
because $\widehat \rho^{\,\mathrm{SM}} = \widehat \rho^{\, \mathrm{NPM}} = 1$.

Since $m_Z$ is one of the input observables
and since $\bar c^2$ is defined
in terms of the input observables through Eq.~\eqref{eq:s2c2 nohat},
$\bar \rho^\mathrm{M}$ in Eq.~\eqref{eq:s2c2 definition}
is just equivalent to $\left( m_W^2 \right)^\mathrm{M}$.
Therefore,
Eq.~\eqref{eq:rhobar SM-NPM oblique} may be rewritten as
\be
m_W^{\, \mathrm{NPM}} = m_W^{\, \mathrm{SM}}
\left[ 1 + \frac{\bar c^2}{2 \left( \bar c^2 - \bar s^2 \right)}\, \alpha T
+ \frac{\alpha S}{4 \left( \bar s^2 - \bar c^2 \right)}
+ \frac{\alpha U}{8 \bar s^2} \right].
\label{eq:mw NPM vs SM}
\ee
Equation~\eqref{eq:mw NPM vs SM} has been used by many authors
to evaluate $m_W$ in their favorite NP models.

If one has a model M with $\widehat\rho\neq1$,
then plugging Eqs.~\eqref{eq:all inputs} into Eq.~\eqref{eq:s2 definition}
one gets
\begin{equation}
  \rho = \widehat \rho^{\,\mathrm{M}}
  \left( 1 + \alpha T^\mathrm{M}  - \alpha K^\mathrm{M}
  - \frac{s^2}{c^2}\, \delta_{Gc}^\mathrm{M} \right)
  \label{eq:rho corrections}
\end{equation}
instead of Eq.~\eqref{eq:rhobar SM},
\cf\ Eq.~\eqref{76788549}.
Note that in Eq.~\eqref{eq:s2 definition}
$\rho$ is just a function of the input,
\cf\ Eq.~\eqref{vvofpd}.
Thus,
$\rho$ is model-independent,
\viz\ it is the same in the BM and in an extension thereof.
Hence,
dividing Eq.~\eqref{eq:rho corrections} in a BBM by the same equation in the BM
and using the assumption in Eq.~\eqref{za1},
one obtains
\begin{equation}
  \label{cjgodpa}
  1 = \frac{\widehat \rho^{\, \mathrm{BBM}}}{\widehat \rho^{\, \mathrm{BM}}}
  \left( 1 + \alpha T  - \alpha K \right).
\end{equation}
Therefore,
\be
\frac{\widehat \rho^{\,\mathrm{BBM}}}{\widehat \rho^{\,\mathrm{BM}}}
=  1 - \alpha T + \alpha K.
\label{eq:rho BM vs BBM}
\ee

On its own,
Eq.~\eqref{eq:rho BM vs BBM} does not have much physical meaning,
because
$T$,
$\widehat \rho^{\,\mathrm{BBM}}$,
and $\widehat \rho^{\,\mathrm{BM}}$ all contain UV divergences.
However,
as we will shortly see,
Eq.~\eqref{eq:rho BM vs BBM} is crucial,
because it allows us to eliminate the divergent $T$ from observable quantities.

\subsection{The parameter \texorpdfstring{$\rho_\star$}{rho star}
  and the substitution \texorpdfstring{$T \to K$}{ T to K}}

The parameter $\rho_\star$ is defined to be
the ratio between the charged and the neutral
effective Fermi constants~\cite{peskin1992}:
\be
\label{rhostar}
\rho^\mathrm{M}_\star \equiv \frac{G^\mathrm{M}_{F(\mathrm{neutral})}}{G_{F(\mathrm{charged})}},
\ee
where $G^\mathrm{M}_{F(\mathrm{neutral})}$ is a prediction for a model M for low-energy neutrino scattering
via an intermediate $Z^0$. 
The relation between the bare and the one-loop charged Fermi constant
is given by Eq.~\eqref{eq:GF charged input}.
The neutral Fermi constant is similarly expressed as
\be
\label{eq:neutral Fermi}
G^\mathrm{M}_{F(\mathrm{neutral})} = \widehat G^\mathrm{M}_{F(\mathrm{neutral})}
\left[ 1 - \frac{\Pi^\mathrm{M}_{ZZ} \left( 0 \right)}{m_Z^2}
  + \frac{2}{sc}\, \frac{\Pi^\mathrm{M}_{ZA} \left( 0 \right)}{m_Z^2}
  + \delta^\mathrm{M}_{Gn} \right],
\ee
where $\delta^\mathrm{M}_{Gn}$ stands for the vertex corrections,
fermion self-energy corrections,
and box corrections in neutral-current neutrino scattering
for a given model M.
Using
Eq.~\eqref{eq:neutral Fermi}
together with Eq.~\eqref{eq:GF charged input} and
\bs
\ba
\widehat G^\mathrm{M}_{F(\mathrm{charged})}
&=& \frac{\pi\, \widehat \alpha^\mathrm{M}}{\sqrt{2}\,
  \left( \widehat s^{\, 2} \right)^\mathrm{M}
  \left( \widehat m_W^{\, 2} \right)^\mathrm{M}},
\\
\widehat G^\mathrm{M}_{F(\mathrm{neutral})}
&=& \frac{\pi\, \widehat \alpha^{\mathrm{M}}}{\sqrt{2}\,
    \left( \widehat s^{\, 2} \right)^\mathrm{M}
    \left( \widehat c^{\, 2} \right)^\mathrm{M}
    \left( \widehat m_Z^{\, 2}\right)^\mathrm{M}},
\\
\widehat \rho^\mathrm{M} &=& \frac{\left( \widehat m_W^{\, 2}
  \right)^\mathrm{M}}{\left( \widehat c^{\, 2} \right)^\mathrm{M}
  \left( \widehat m_Z^{\, 2} \right)^\mathrm{M}},
\ea
\es
we get
\be
\rho_\star^\mathrm{M} = \widehat \rho^{\,\mathrm{M}}\,
\left( 1 + \alpha T^\mathrm{M}
  + \delta_{Gn}^\mathrm{M} - \delta_{Gc}^\mathrm{M} \right).
\label{eq:rhostar initial}
\ee
Using $\widehat \rho^{\,\mathrm{M}}=1$ in Eq.~\eqref{eq:rhostar initial}
for both the SM and an NPM,
and assuming that
\be
\delta_{Gn}^\mathrm{M^\prime} - \delta_{Gn}^\mathrm{M}
\ \mathrm{is\ negligible},
\label{za2}
\ee
just as Eq.~\eqref{za1},
one obtains~\cite{peskin1992}
\be
\frac{\rho_\star^\mathrm{NPM}}{\rho_\star^\mathrm{SM}} = 1 + \alpha T\, ,
\label{rhostarSM}
\ee

In a BM and a BBM,
one must leave $\widehat \rho^{\,\mathrm{M}}$ free
in Eq.~\eqref{eq:rhostar initial},
which then yields
\begin{equation}
  \frac{\rho_\star^\mathrm{BBM}}{\rho_\star^\mathrm{BM}} =
  \frac{\widehat \rho^{\, \mathrm{BBM}}}{\widehat \rho^{\, \mathrm{BM}}}
  \left( 1 + \alpha T \right). \label{eq:rhostar BBM with rhohat}
\end{equation}
Plugging
$\widehat \rho^{\,\mathrm{BBM}} \! \left/ \widehat \rho^{\, \mathrm{BM}} \right.$
from Eq.~\eqref{eq:rho BM vs BBM}
into Eq.~\eqref{eq:rhostar BBM with rhohat} one gets
\begin{equation}
    \frac{\rho_\star^\mathrm{BBM}}{\rho_\star^\mathrm{BM}} = 1 + \alpha K.
    \label{jvuf99}
\end{equation}
Equation~\eqref{jvuf99} is gauge-invariant,
finite,
and is written only in terms of observable quantities and bare self-energies.
By using Eq.~\eqref{eq:rho BM vs BBM} to get rid of the dependence on
$\widehat \rho^{\,\mathrm{BBM}} \! \left/ \widehat \rho^{\,\mathrm{BM}} \right.$,
we effectively replaced the divergent quantity $T$
by the finite quantity $K$,
\cf\ Eqs.~\eqref{rhostarSM} and~\eqref{jvuf99}.
This is a \emph{general substitution rule}
that holds for all the EW observables
when one uses $m_W$ as an input instead of $\widehat \rho = 1$.\footnote{Indeed,
one may trade the oblique parameter {$T$} for {$m_W$}
in models with {$\widehat \rho = 1$} too. 
However,
this is not really practical,
because instead of calculating  $T$ one would then have to calculate
the difference in the predictions for $m_W$ between the SM and an NPM.}

\subsection{Summary}

The main  difference between the cases ``$\widehat \rho = 1$''
and ``free $\widehat \rho$'' is the number of input observables.
When $\widehat \rho$ is free,
one uses the additional input observable $m_W$
to cancel the appearances of $\widehat \rho$ in every other observable
via Eq.~\eqref{eq:rho BM vs BBM}. 
By doing this one effectively replaces
the parameter $T$
by a linear combination of the parameters $S$ and $U$
\be
\label{eq:T subs}
T \ \to \ K \equiv \frac{S}{2 c^2} + \frac{s^2 - c^2}{4 s^2 c^2}\,
U.
\ee
%
One may use the substitution of Eq.~\eqref{eq:T subs}
on the expressions for all the observables,
listed in Appendix~\ref{app:observable list},
which hold for the comparison between an NPM and the SM, 
to arrive at analogous expressions for the comparison
between a BBM and its BM.


Having this substitution rule,
one must also replace $\bar{s}$ and $\bar{c}$ in Eq.~\eqref{eq:s2c2 nohat}
by $s$ and $c$ in Eq.~\eqref{eq:s2 nohat},
\ie
\begin{equation}\label{eq: trig sub}
    \bar{s}\to s, \quad \bar{c}\to c.
\end{equation}
For example,
taking values from~\cite{particledatagroup2022} one has $\bar{s}^2=0.233563$,
while either $s^2=0.230403$ or $s^2=0.230079$
depending on whether $m_W$ is taken from the PDG~\cite{particledatagroup2022}
or from the recent CDF result~\cite{cdfcollaboration+++2022},
respectively. 

The definitions of the OPs stay the same in terms of the self-energies,
but they must be computed without ever assuming $m_W= m_Z \cos \theta_W$,
in models where $\widehat \rho$ is not fixed.

\section{Extensions of the SM with scalar triplets}
\label{sec:example}

In the following,
we give an explicit calculation of the OPs
in a case where both the BM and the BBM have $\hat \rho \neq 1$.
We choose to consider extensions of the SM by scalar triplets,
both because of their relative simplicity
and because such models may be phenomenologically important
to interpret the CDF measurement of $m_W$
mentioned in Section~\ref{sec:intro}.
In addition,
the 95~GeV excess measured by the
CMS Collaboration~\cite{sirunyan2019,thecmscollaboration2023,gascon-shotkinsusan2023,thecmscollaboration2023a}
could be accounted for by a model with a triplet,
although this possibility
has not yet been thoroughly explored~\cite{ashanujjaman2023}.

\subsection{The BBM and the BM}\label{sec:BBM model}

We consider an $SU(2) \times U(1)$ gauge model
with the following scalar multiplets of $SU(2)$:
\begin{itemize}
\item One doublet with $Y_H = 1/2$,
  \textit{viz}.
  \be
  \Phi = \left( \begin{array}{c} \phi^+ \\ \phi \end{array} \right),
  \qquad \phi = \widehat v + \frac{R_1 + i I_1}{\sqrt{2}},
  \label{Phi}
  \ee
where $\widehat v$ is a real and positive VEV.
\item One triplet with $Y_H = 1$,
  \textit{viz}.
  \be
  \Theta = \left( \begin{array}{c} \theta^{++} \\ \theta^+ \\ \theta
  \end{array} \right),
  \qquad \theta = \widehat t + \frac{R_2 + i I_2}{\sqrt{2}},
  \label{Theta}
  \ee
where $\widehat t$ is a real and positive VEV.
  The charge-2 field $\theta^{++}$ is an eigenstate of mass
  with mass $m_{++}$.
\item One real triplet with $Y_H = 0$,
  \textit{viz}.
  \be
  \Delta = \left( \begin{array}{c} \delta^+ \\ \delta \\ - \delta^-
  \end{array} \right),
  \qquad \delta = \widehat u + R_3,
  \label{Delta}
  \ee
  where $\delta^- = {\delta^+}^\ast$,
  $\delta = \delta^\ast$,
 and $\widehat u$ is a real and positive VEV.
\end{itemize}
In Eqs.~\eqref{Phi}--\eqref{Delta},
the fields $R_1$, $R_2$, $R_3$, $I_1$, and $I_2$ are real.

The gauge-boson masses are
\bs \label{eq:mw and mz in bbm}
\ba
\widehat m_W &=& \frac{\widehat g}{\sqrt{2}}\, \widehat{\varrho}_W,
 \quad
\widehat{\varrho}_W \equiv \sqrt{\widehat v^2 + 2 \widehat t^2 + 2 \widehat u^2},
\\
\widehat m_Z &=& \frac{\widehat g}{\sqrt{2}\, \widehat c}\, \widehat{\varrho}_Z,
 \quad
\widehat{\varrho}_Z \equiv \sqrt{\widehat{v}^{2} + 4 \widehat{t}^{2}}.
\ea
\es
This leads to 
\be
\frac{1}{2} \le
\widehat \rho = \frac{\widehat{\varrho}^{\ 2}_W}{\widehat{\varrho}^{\ 2}_Z} \le \infty,
\label{eq:rho BBM}
\ee
\viz\ $\widehat \rho$ is free.

The mixing of the charge-1 scalars is given by
\be \label{eq:mix charged}
\left( \begin{array}{c} G^+ \\ H_2^+ \\ H_3^+ \end{array} \right)
= \mathcal{O}
\left( \begin{array}{c} \phi^+ \\ \delta^+ \\ \theta^+ \end{array} \right),
\ee
where $G^+$ is the charged Goldstone boson,
the $H_j^+$ ($j = 2, 3$) are physical charged scalars with masses $m_{j+}$,
and
\be
\mathcal{O} = \left( \begin{array}{ccc} \widehat v \left/ \widehat{\varrho}_W \right. &
  \sqrt{2} \, \widehat u \left/ \widehat{\varrho}_W \right. &
  \sqrt{2} \, \widehat t \left/ \widehat{\varrho}_W \right. \\
  \mathcal{O}_{21} & \mathcal{O}_{22} & \mathcal{O}_{23} \\
  \mathcal{O}_{31} & \mathcal{O}_{32} & \mathcal{O}_{33}
\end{array} \right)
\ee
is a $3 \times 3$ orthogonal matrix.
The orthogonality relations
\be
\sum_{k=1}^3 \mathcal{O}_{mk}\mathcal{O}_{nk} = \delta_{mn},
\ee
together with the fixed first row,
leave only one free parameter in $\mathcal{O}$.

The mixing of the pseudoscalars is given by
\be \label{eq:mix pesudoscalars}
\left( \begin{array}{c} G \\ A \end{array} \right)
= \frac{1}{\widehat{\varrho}_Z} \left( \begin{array}{cc} \widehat v & 2 \widehat t \\ - 2 \widehat t & \widehat v
\end{array} \right) \left( \begin{array}{c} I_1 \\ I_2 \end{array} \right),
\ee
where $G$ is the neutral Goldstone boson
and $A$ is a physical pseudoscalar with mass $m_A$.

The mixing of the neutral scalars is given by
\be
\left( \begin{array}{c} S_1 \\ S_2 \\ S_3 \end{array} \right)
= O
\left( \begin{array}{c} R_1 \\ R_3 \\ R_2 \end{array} \right),
\ee
where $O$ is a $3 \times 3$ real orthogonal matrix
and the $S_k$ ($k = 1, 2, 3$) are physical scalars with mass $m_k$.

We take the model exposed above to be the BBM.
The BM may then be
\begin{itemize}
\item either identical to the BBM without $\Delta$,
  which we call BM1;
\item or identical to the BBM without $\Theta$,
  and we call this BM2.
\end{itemize}
Both the BM1 and the BM2 have free $\widehat \rho$
and therefore their predictions may be compared
with those of the BBM through OPs.
One should however note that the BM1 has $\widehat u=0$,
hence $\widehat \rho < 1$,
while the BM2 has $\widehat t=0$,
hence $\widehat \rho > 1$,
and therefore one may compare each of them to the BBM
only for well-defined (and non-intersecting) ranges
of $\widehat \rho^{\, \mathrm{BBM}}$.
One must also take into consideration that,
while the input values of $m_W$ and $m_Z$ are the same
for both the BBM and the BMs,
the values of $v$,
$t$,
and $u$ will turn out different in the BBM and the corresponding BM.

\subsection{Oblique parameters}

In general, 
one may write each model parameter $O^\mathrm{M}$
(M$\, =\, $BBM, BM1, BM2)
in the form
\be
O^\mathrm{M} =
O^\mathrm{M}_\mathrm{scalars}
+ O^\mathrm{M}_\mathrm{fermions}
+ O^\mathrm{M}_\mathrm{gauge},
\ee
where
\begin{itemize}
\item $O^\mathrm{M}_\mathrm{scalars}$ is the contribution to $O^\mathrm{M}$
  from diagrams that have \emph{at least one physical} scalar in the loop
  (hence,
  one includes diagrams where either a gauge or a Goldstone boson
  is in one internal line of the loop
  and a physical scalar is in the other internal line).
\item $O^\mathrm{M}_\mathrm{fermions}$ is the contribution to $O^\mathrm{M}$
  from diagrams that have two fermions in the loop;
\item $O^\mathrm{M}_\mathrm{gauge}$ is the contribution to $O^\mathrm{M}$
  from diagrams that have \emph{only} gauge bosons,
  Goldstone bosons,
  and/or ghosts in the loop.
\end{itemize}
$O^\mathrm{M}_\mathrm{fermions}$ is gauge-independent because
it has neither gauge nor Goldstone bosons in the loops.
We have found that $O^\mathrm{M}_\mathrm{scalars}$ is also gauge-independent.
On the other hand,
$O^\mathrm{M}_\mathrm{gauge}$ depends on the gauge,
and that gauge-dependence only cancels with the analogous contribution in the BM
upon subtraction. 
This means that the gauge-dependence of $O^\mathrm{M}_\mathrm{gauge}$
is the same for BBM,
BM1,
and BM2 provided the same value of $\rho$ is utilized
and the input observables in the renormalization procedure
of the EW sector are the same.
Thus,
gauge-dependence cancels out in the subtracted OPs,
\viz\ in $O^\mathrm{BBM}_\mathrm{gauge} - O^\mathrm{BM}_\mathrm{gauge}$.
We have explicitly checked this fact
for the triplet extensions of the SM introduced above.

The contribution $O^\mathrm{M}_\mathrm{scalars}$
is not necessarily finite
before the subtraction of the same quantity in the base model.
Indeed, scalar contributions to
both the $S$ and $U$ parameters are UV divergent.
However,
their divergences depend only
on the \emph{two} quantities $\varrho_W$ and $\varrho_Z$---which are
functions of the input observables---instead of depending
on all \emph{three} VEVs $v$,
$t$,
and $u$.
In turn,
the UV divergences cancel out upon subtraction. 
On the other hand,
the parameters $V^\mathrm{M}_\mathrm{scalars}$,
$W^\mathrm{M}_\mathrm{scalars}$,
and $X^\mathrm{M}_\mathrm{scalars}$ are finite without the need of a subtraction.

It is interesting to note that in a gauge
where all the gauge parameters are equal,
\ie\ $\xi_W=\xi_Z=\xi_A \equiv \xi$,
the UV divergences of $S^\mathrm{M}_\mathrm{scalars}$
and $U^\mathrm{M}_\mathrm{scalars}$
exactly cancel out the UV divergences appearing in
$S^\mathrm{M}_\mathrm{gauge}$ and $U^\mathrm{M}_\mathrm{gauge}$,
respectively
(see Appendix~\ref{app:divergences} for an explanation of this fact),
leading to finite model parameters.

In the following,
we give analytical expressions for all the $O^\mathrm{BBM}_\mathrm{scalars}$.
The model parameters for the BM1 and the BM2 may be recovered
by setting appropriate VEVs to zero
and by removing the contributions from particles
that do not exist in those base models.
The parameters $O^\mathrm{BBM}$ are expressed
in terms of Passarino--Veltman functions defined in Appendix~\ref{app:PaVe}.
Finally,
note that
\be
S^\mathrm{M} + U^\mathrm{M} = \frac{4 s^2}{\alpha} \left[
\widetilde{\Pi}^\mathrm{M}_{WW} \left( m_W^2 \right)
- \frac{c}{s}\, \Pi_{ZA}^{\prime\, \mathrm{M}} \left( 0 \right)
- \Pi_{AA}^{\prime\, \mathrm{M}} \left( 0 \right) \right]
\ee
is usually easier to compute than $U^\mathrm{M}$ itself.
For this reason,
we write firstly $S^\mathrm{BBM}$ and then
$U^\mathrm{BBM} = - S^\mathrm{BBM} + \left( S^\mathrm{BBM} + U^\mathrm{BBM} \right)$
instead of writing the full expression for $U^\mathrm{BBM}$.

The BBM model parameters are as follows
\bs
\ba
S^\mathrm{BBM}_\mathrm{scalars} &=&
\frac{1}{\pi} \left\{ \vphantom{\sum_{j=2}^3}
4 \left( s^2 - c^2 \right)^2
\bar B_{00} \left( m^2_Z, m^2_{++}, m^2_{++} \right)
\right.
\\ & &
- 8 \left( s^4 + c^4 \right)
B^\prime_{00} \left( 0, m^2_{++}, m^2_{++} \right)
\\ & &
+ 4\, \sum_{k=1}^3
\frac{\left( v O_{k3} - t O_{k1} \right)^2}{\varrho_Z^2}\,
\bar B_{00} \left( m^2_Z, m^2_A, m^2_k \right)
\\ & &
+ \sum_{k=1}^3
\frac{\left( v O_{k1} + 4 t O_{k3} \right)^2}{\varrho_Z^2}\ m^2_Z\,
\bar F \left( m^2_Z, m^2_Z, m^2_k \right)
\\ & &
+ 4\, \sum_{j=2}^3
\frac{\left( u \mathcal{O}_{j2} - t \mathcal{O}_{j3}
  \right)^2}{\varrho_W^2}\ m^2_W\,
\bar F \left( m^2_Z, m^2_W, m^2_{j+} \right)
\\ & &
+ 2 \left( \mathcal{O}_{22} \mathcal{O}_{32}
- \mathcal{O}_{23} \mathcal{O}_{33} \right)^2
\bar B_{00} \left( m^2_Z, m^2_{2+}, m^2_{3+} \right)
\\ & &
+ \sum_{j=2}^3
\left( s^2 - c^2 + \mathcal{O}_{j3}^2 - \mathcal{O}_{j2}^2 \right)^2
\bar B_{00} \left( m^2_Z, m^2_{j+}, m^2_{j+} \right)
\\ & & \left.
- 2\sum_{j=2}^3 \left[ (s^4 + c^4) + \left( s^2 - c^2 \right)
  \left( \mathcal{O}_{j3}^2 - \mathcal{O}_{j2}^2 \right) \right]
B_{00}^\prime \left( 0, m^2_{j+}, m^2_{j+} \right) 
\right\}; 
\ea
\es
\bs
\ba
U^\mathrm{BBM}_\mathrm{scalars} &=& - S^\mathrm{BBM}_\mathrm{scalars} + 
\frac{1}{\pi} \left\{
\vphantom{+ 2 \sum_{j=2}^3
\frac{\left( u \mathcal{O}_{j2} - t \mathcal{O}_{j3} \right)^2}{\varrho_Z^2}}
- 8 B^\prime_{00} \left( 0, m_{++}^2, m_{++}^2 \right)
\right. 
\\ & &
+ 2 \sum_{j=2}^3 \left( \mathcal{O}_{j3}^2 - \mathcal{O}_{j2}^2 - 1 \right)
B^\prime_{00} \left( 0, m_{j+}^2, m_{j+}^2 \right)
\\ & &
+ 4 \sum_{j=2}^3 \mathcal{O}_{j3}^2
\bar B_{00} \left( m_W^2, m_{++}^2, m_{j+}^2 \right)
\\ & &
+ 2\, \sum_{j=2}^3 \frac{\left( v \mathcal{O}_{j3}
  - \sqrt{2} t \mathcal{O}_{j1} \right)^2}{\varrho_Z^2}\,
\bar B_{00} \left( m_W^2, m_A^2, m_{j+}^2 \right)
\\ & &
+ \sum_{k=1}^3 \sum_{j=2}^3
\left( O_{k1} \mathcal{O}_{j1}
+ \sqrt{2}\, O_{k3} \mathcal{O}_{j3} + 2\, O_{k2} \mathcal{O}_{j2} \right)^2
\bar B_{00} \left( m_W^2, m_k^2, m_{j+}^2 \right)
\\ & &
+ \frac{8 t^2}{\varrho_W^2}\,
m_W^2\, \bar F \left( m_W^2, m_W^2, m_{++}^2 \right) \hspace{1cm}
\\ & &
+ \sum_{k=1}^3  \frac{\left( v O_{k1} + 2 t O_{k3}
  + 2 \sqrt{2} u O_{k2} \right)^2}{\varrho_W^2}\
m_W^2\, \bar F \left( m_W^2, m_W^2, m_k^2 \right)
\\ & & \left.
+ 2 \sum_{j=2}^3
\frac{\left( u \mathcal{O}_{j2} - t \mathcal{O}_{j3} \right)^2}{\varrho_Z^2}\
m^2_Z\, \bar F \left( m^2_W, m^2_Z, m^2_{j+} \right)
\right\};
\ea
\es
\bs
\label{v5}
\ba
V^\mathrm{BBM}_\mathrm{scalars} &=&
\frac{1}{\pi s^2 c^2} \left\{
\vphantom{\frac{\left( \mathcal{O}_{22} \mathcal{O}_{32}
  - \mathcal{O}_{23} \mathcal{O}_{33} \right)^2}{2}}
\left( s^2 - c^2 \right)^2\,
\widetilde B_{00} \left( m_Z^2, m_{++}^2, m_{++}^2 \right)
\right. \\ & &
+ \sum_{k=1}^3
\frac{\left( v O_{k3} - t O_{k2} \right)^2}{\varrho_Z^2}\
\widetilde B_{00} \left( m_Z^2, m_A^2, m_k^2 \right)
\\ & &
+ \frac{m_Z^2}{4}\, \sum_{k=1}^3
\frac{\left( v O_{k2} + 4 t O_{k3} \right)^2}{\varrho_Z^2}\
\widetilde F \left( m_Z^2, m_Z^2, m_k^2 \right)
\\ & &
+ m_W^2 \sum_{j=2}^3
\frac{\left( u \mathcal{O}_{j2} - t \mathcal{O}_{j3} \right)^2}{\varrho_W^2}\
\widetilde F \left( m_Z^2, m_W^2, m_{j+}^2 \right)
\\ & &
+ \sum_{j=2}^3
\frac{\left( s^2 - c^2 + \mathcal{O}_{j3}^2 - \mathcal{O}_{j2}^2 \right)^2}{4}\
\widetilde B_{00} \left( m_Z^2, m_{j+}^2, m_{j+}^2 \right)
\\ & & \left.
+ \frac{\left( \mathcal{O}_{22} \mathcal{O}_{32}
  - \mathcal{O}_{23} \mathcal{O}_{33} \right)^2}{2}\
\widetilde B_{00} \left( m_Z^2, m_{2+}^2, m_{3+}^2 \right)
\right\};
\ea
\es
\bs
\label{w5}
\ba
W^\mathrm{BBM}_\mathrm{scalars} &=&
\frac{1}{\pi s^2} \left\{
\sum_{j=2}^3 \mathcal{O}_{j3}^2\,
\widetilde B_{00} \left( m_W^2, m_{++}^2, m_{j+}^2 \right)
\right. \\ & &
+ \sum_{j=2}^3 \frac{\left( v \mathcal{O}_{j3} - \sqrt{2} t \mathcal{O}_{j1}
  \right)^2}{2 \varrho_Z^2}\
\widetilde B_{00} \left( m_W^2, m_A^2, m_{j+}^2 \right)
\\ & &
+ \sum_{k=1}^3 \sum_{j=2}^3
\frac{\left( O_{k1} \mathcal{O}_{j1} + \sqrt{2}\, O_{k3} \mathcal{O}_{j3}
+ 2\, O_{k2} \mathcal{O}_{j2} \right)^2}{4}\
\widetilde B_{00} \left( m_W^2, m_k^2, m_{j+}^2 \right)
\\ & &
+ \frac{2 m_W^2 t^2}{\varrho_W^2}\
\widetilde F \left( m_W^2, m_W^2, m_{++}^2 \right)
\\ & &
+ \frac{m_W^2}{4}\, \sum_{k=1}^3
\frac{\left( v O_{k1} + 2 t O_{k3} + 2 \sqrt{2} u O_{k2} \right)^2}{\varrho_W^2}\
\widetilde F \left( m_W^2, m_W^2, m_k^2 \right)
\\ & & \left.
+ \frac{m_Z^2}{2}\, \sum_{j=2}^3
\frac{\left( u \mathcal{O}_{j2} - t \mathcal{O}_{j3} \right)^2}{\varrho_Z^2}\
\widetilde F \left( m_W^2, m_Z^2, m_{j+}^2 \right)
\right\};
\ea
\es
\bs
\label{x5}
\ba
X^\mathrm{BBM}_\mathrm{scalars} &=&
\frac{2 \left( s^2 - c^2 \right)}{\pi}
\left[ B^\prime_{00} \left( 0, m_{++}^2, m_{++}^2 \right)
  - \bar B_{00} \left( m_Z^2, m_{++}^2, m_{++}^2 \right) \right]
\\ & &
+ \sum_{j=2}^3 \frac{s^2 - c^2
+ \mathcal{O}_{j3}^2 - \mathcal{O}_{j2}^2}{2 \pi} \left[
  B^\prime_{00} \left( 0, m_{j+}^2, m_{j+}^2 \right)
  - \bar B_{00} \left( m_Z^2, m_{j+}^2, m_{j+}^2 \right) \right].
\hspace*{7mm}
\ea
\es

\section{Oblique corrections to the SM in a model
  with \texorpdfstring{$\widehat \rho \neq 1$}{rho hat neq 1}}
\label{sec4}

When one compares the SM to a model M where $\widehat \rho \neq 1$,
one faces complications due to the renormalization of the EW sector
requiring different numbers of input observables in the two models. 
This mismatch causes problems when one attempts to use
an input observable like $m_W$ in M. 
Indeed,
in the SM $m_W$ has a predicted value $m_W^\mathrm{SM}$.  
If one uses $m_W \neq m_W^\mathrm{SM}$ as an input of M,
then the gauge dependences in the subtracted OPs do not cancel out,
and this makes it necessary to take into account the non-oblique contributions. 
More specifically,
if one takes,
for instance,
$O^\mathrm{BBM}_\mathrm{gauge}$ from the previous section
and the analogous $O^\mathrm{SM}_\mathrm{gauge}$,
they are not equal and therefore the subtraction of one from the other
gives a gauge-dependent result.
This means that some OPs $O=O^\mathrm{M}-O^\mathrm{SM}$
are gauge-dependent and cannot be used to parametrize observables.
This is in addition to the UV-divergent $T$ parameter.

Kennedy and Lynn~\cite{kennedy1989,lynn1989,lynn1992}
have discussed the issue of gauge dependence
in the context of oblique corrections.
They have proposed a method to define gauge-invariant self-energies
by including both vertex and box `universal' contributions in the propagators. 
This approach may allow to define both the (un-subtracted) model parameters
and the (subtracted) OPs in a gauge-independent way.
However,
it is unclear whether the increased complexity
of such generalized parameters would justify the loss of accuracy,
when compared to a full calculation.

It remains an open question whether it is at all possible
to construct a convenient parametrization,
similar to the OPs,
when comparing a model with $\widehat \rho = 1$
to a model with a free $\widehat \rho$.

\section{Conclusions}
\label{sec:conclusions}
The predictive power of models with a free $\widehat \rho$
is smaller than the one of models where $\widehat \rho = 1$,
since in the former case one needs to remove the additional freedom
by providing information about an additional observable.
When $\widehat \rho$ is free,
the oblique parameter $T$ is UV-divergent
and cannot be used to parametrize UV-finite observables.\footnote{Note that,
as a matter of fact,
$T$ is present in the predictions for \emph{all} the EW observables
in an NPM in tables~\ref{tab:Z decays},
\ref{tab:other obs},
and~\ref{tab:Z-pole}.}

We have chosen to use the mass of the gauge bosons $W^\pm$ as additional input,
along with the three usual quantities:
the fine-structure constant,
the Fermi constant,
and the mass of the $Z^0$.
With these four inputs,
the previously divergent $T$ is simply discarded
and only five oblique parameters remain---$S$,
$U$,
$V$,
$W$,
and $X$.
It can be said that the UV-divergent parameter had to be renormalized,
and that the renormalization was performed by fixing the mass of the $W^\pm$
to its experimental value.
Quite conveniently,
the parametrization of observables
in terms of the five remaining oblique parameters
may be reached from the six-parameter case via a simple substitution
of $T$ by a linear combination of $S$ and $U$,
\textit{cf.}\ Eq.~\eqref{eq:T subs}.

Our approach provides a genuine oblique parameter formalism,
wherein all the oblique parameters are UV-finite and gauge-independent.
The downside of our approach is that it is applicable only
when both models being compared have broken custodial symmetry,
\ie\ free $\widehat{\rho}$.
If one of the models being compared has custodial symmetry
and the other one does not,
the parameter $T$ is UV-divergent
and other oblique parameters are either gauge-dependent or divergent.
These obstacles \emph{are} informative,
as they tell us that one should not use
the oblique parameter formalism to compare a model that has custodial symmetry
to another model that does not have it.
Although this no-go is not new in the literature,
our paper should serve as a useful reminder,
in light of the increased popularity of custodial symmetry-violating models,
especially models with triplet scalars.

In Section~\ref{sec:example} we have presented analytical expressions
for the five remaining oblique parameters
in a two-scalar-triplet extension of the SM,
where both triplets---one with hypercharge 1
and the other one with zero hypercharge---have nonzero
vacuum expectation values.
The appendices contain parametrizations of observables
for pairs of models with or without custodial symmetry,
which should serve as a convenient reference.

\vspace*{5mm}
\paragraph{Acknowledgments}
The authors thank Duarte Fontes
for reading and commenting on a draft of this paper.
S.D.\ and V.D.\ thank the Lithuanian Academy of Sciences
for funding received through the project DaFi2021.
L.L.\ thanks the Portuguese Foundation for Science and Technology
for support through the projects UIDB/00777/2020,
UIDP/00777/2020,
CERN/FIS-PAR/0002/2021,
and CERN/FIS-PAR/0019/2021.

\begin{appendix}
\section{List of observables for SM \vs\ NPM} \label{app:observable list}

In Tables~\ref{tab:Z-pole},
\ref{tab:Z decays},
and~\ref{tab:other obs} we list the expressions for the oblique corrections
of various EW observables,
in the case of a New Physics Model (NPM)
that is an extension of the Standard Model (SM),
both models featuring $m_W = m_Z \cos{\theta_W}$ at tree-level.

Note that some of the observables in Table~\ref{tab:Z-pole},
\viz\ $\sigma_\mathrm{had}$,
$R_\ell$,
$R_b$,
and $R_c$,
are derived from observables in Table~\ref{tab:Z decays}.
Also note that some other observables in Table~\ref{tab:Z-pole},
\viz\ $A^\ell_\mathrm{FB}$, $A^b_\mathrm{FB}$, and $A^c_\mathrm{FB}$,
are derived from other observables in the same Table~\ref{tab:Z-pole}.

The oblique corrections in the tables
are calculated from the tree-level expressions,
including the residues of the fields in the case of decaying particles.
The corresponding expressions are listed
in the first columns of the tables.
There,
we have used the definitions
\bs
\ba
s^2_0 &\equiv& s^2_\mathrm{effective} \left( p^2 \approx 0 \right),
\\
s^2_Z &\equiv& s^2_\mathrm{effective} \left( p^2 \approx m_Z^2 \right),
\ea
\es
where $s^2_\mathrm{effective}$ is the effective sine-squared
of the Weinberg angle~\cite{altarelli1989}.
Note that some of the parameters in these expressions are bare
and are denoted by hats;
they must be related to the input observables via the self-energies.
On the other hand,
the masses of decaying particles,
\viz\ $m_W$ and $m_Z$,
are pole masses and thus are written without hats.  

The expressions in the right columns of the tables
are derived by performing the following substitutions
in the left columns:
\bs
\label{mvklof}
\ba
\widehat c\, \widehat m_Z & \to &
m_W \left[ 1
+ \frac{\alpha S}{4 \left( \bar s^2 - \bar c^2 \right)}
+ \frac{\bar c^2}{2 \left( \bar c^2 - \bar s^2 \right)}\, \alpha T
+ \frac{\alpha U}{8 \bar s^2} \right],
\label{eq:cmz replacement }
\\
s_0^2 &\to& \bar s^2 \left( 1 + \frac{\alpha \bar F_0}{4 \bar s^2} \right),
\label{eq:s0 replacement}
\\
s_Z^2 &\to& \bar s^2 \left( 1 + \frac{\alpha \bar F_Z}{4 \bar s^2} \right),
\label{eq:sz replacement}
\\
\frac{\widehat \alpha}{\widehat s^{2}}
\left[ 1 + \Pi_{WW}^{\prime} \left( m_W^2 \right) \right]
& \to &  \frac{\alpha}{ \bar s^{2}}
\left[ 1 + \frac{\alpha S}{2 \left( \bar s^2 - \bar c^2 \right)}
+ \frac{\bar c^2}{\bar c^2 - \bar s^2}\, \alpha T
+ \frac{\alpha U}{4 \bar s^2} + \alpha W \right],
\label{eq:W decay replacement}
\\ 
\frac{ \widehat \alpha }{\widehat c^2\, \widehat s^2}
\left[ 1 + \Pi_{ZZ}^\prime \left( m_Z^2 \right) \right] &\to&
\frac{\alpha}{\bar c^2 \bar s^2} \left( 1 + \alpha T + \alpha V \right),
\label{eq:Z decay replacement}
\\
\rho_\star &\to& 1 + \alpha T,
\label{eq:rhostar replacement}
\ea
\es
where we have abbreviated
the oblique corrections to the effective Weinberg angle as
\bs
\label{eq:Fz and F0}
\ba
\bar F_0 &\equiv& \frac{S}{\bar c^2 - \bar s^2}
+ \frac{4 \bar s^2 \bar c^2}{\bar s^2 - \bar c^2}\, T
\label{fff000}
\\
\bar F_Z &\equiv& \bar F_0 + 4 X.
\label{fffzzz}
\ea
\es
Making these replacements,
normalizing to the SM tree-level expressions,
and expanding to first order in the OPs
leads to the expressions shown in the right columns of the tables.

\renewcommand\arraystretch{1.75}
\begin{table}[ht]
\center
\begin{tabular}{|c|c|}
\hline 
Observable $O$ & $\left. O^\mathrm{NPM} \right/ O^\mathrm{SM} - 1$
\tabularnewline
\hline 
\hline 
$\sigma_{\text{had}} = \frac{12 \pi}{m_Z^2}\,
\frac{\Gamma_{Z\to\ell\bar{\ell}}\,
  \Gamma_{Z\to\text{hadrons}}}{\left( \Gamma_{Z\to\text{all}} \right)^{2}}$ &
$\left( \frac{4 \bar s^2 - 1}{8 \bar s^4 - 4 \bar s^2 + 1}
+ \frac{44 \bar s^2 - 21}{88 \bar s^4 - 84 \bar s^2 + 45}
- \frac{160 \bar s^2 - 60}{160 \bar s^4 - 120 \bar s^2 + 63} \right)
\alpha \bar{F}_Z$
\tabularnewline
\hline 
\multirow{1}{*}{$R_{\ell}
  = \frac{\Gamma_{Z\to\text{hadrons}}}{\Gamma_{Z\to\ell\bar{\ell}}}$} &
$\left( \frac{44 \bar s^2 - 21}{88 \bar s^4 - 84 \bar s^2 + 45}
- \frac{4 \bar s^2 - 1}{8 \bar s^4 - 4 \bar s^2 + 1} \right)
\alpha \bar F_Z$
\tabularnewline
\hline 
$A_{FB}^\ell = \frac{3}{4} \left( A_{LR}^\ell \right)^2$ &
$\left( \frac{2}{4 \bar s^2 - 1}
+ \frac{2 - 8 \bar s^2}{1 - 4 \bar s^2 + 8 \bar s^4} \right)
\alpha \bar F_Z$
\tabularnewline
\hline 
$R_{b}=\frac{\Gamma_{Z\to b\bar{b}}}{\Gamma_{Z\to\text{hadrons}}}$ &
$\left( \frac{4 \bar s^2 - 3}{8 \bar s^4 - 12 \bar s^2 + 9}
- \frac{44 \bar s^2 - 21}{88 \bar s^4 - 84 \bar s^2 + 45} \right)
\alpha \bar F_Z$
\tabularnewline
\hline 
$R_{c}=\frac{\Gamma_{Z\to c\bar{c}}}{\Gamma_{Z\to\text{hadrons}}}$ &
$\left( \frac{16 \bar s^2 - 6}{32 \bar s^4 - 24 \bar s^2 + 9}
- \frac{44 \bar s^2 - 21}{88 \bar s^4 - 84 \bar s^2 + 45} \right)
\alpha \bar F_Z$
\tabularnewline
\hline 
$A_{FB}^b = \frac{3}{4}\, A_{LR}^\ell\, A_{LR}^b$ &
$\left( \frac{1}{4 \bar s^2 - 1}
+ \frac{1}{4 \bar s^2 - 3}
+ \frac{3 - 4 \bar s^2}{9 - 12 \bar s^2 + 8 \bar s^4}
+ \frac{1 - 4 \bar s^2}{1 - 4 \bar s^2 + 8 \bar s^4} \right)
\alpha \bar{F}_Z$
\tabularnewline
\hline 
$A_{FB}^c = \frac{3}{4}\, A_{LR}^\ell\, A_{LR}^c$ &
$\left( \frac{1}{4 \bar s^2 - 1}
+ \frac{2}{8 \bar s^2 - 3}
+ \frac{1 - 4 \bar s^2}{1 - 4 \bar s^2 + 8 \bar s^4}
+ \frac{6 - 16 \bar s^2}{9 - 24 \bar s^2 + 32 \bar s^4}
\right)\alpha \bar{F}_Z$
\tabularnewline
\hline 
$s_Z^2 = \widehat s^2
- \bar s \bar c\, \frac{\Pi_{ZA} \left( m_Z^2 \right)}{m_Z^2}$ &
$\frac{\alpha \bar F_Z}{4 \bar s^2}$\tabularnewline
\hline 
$A_{LR}^{\ell} = 2\, \frac{1 - 4 s_Z^2}{1 + \left( 1 - 4 s_Z^2 \right)^2}$ &
$\left( \frac{1}{4 \bar s^2 - 1}
+ \frac{1 - 4 \bar s^2}{1 - 4 \bar s^2 + 8 \bar s^4} \right)
\alpha \bar F_Z$
\tabularnewline
\hline 
$A_{LR}^b = \frac{\left( - \frac{1}{2} + \frac{1}{3}\, s_Z^2 \right)^2
  - \left( \frac{1}{3}\, s_Z^2 \right)^2}{\left( - \frac{1}{2}
  + \frac{1}{3}\, s_Z^2 \right)^{2} + \left( \frac{1}{3}\, s_Z^2 \right)^2}$ &
$\left( \frac{1}{4 \bar s^2 - 3}
+ \frac{3 - 4 \bar s^2}{9 - 12 \bar s^2 + 8 \bar s^4} \right)
\alpha \bar F_Z$
\tabularnewline
\hline 
$A_{LR}^{c}=\frac{\left( \frac{1}{2} - \frac{2}{3}\, s_Z^2 \right)^2
  - \left( \frac{2}{3}\, s_Z^2 \right)^2}{\left( \frac{1}{2}
  - \frac{2}{3}\, s_Z^2 \right)^2 + \left( \frac{2}{3}\, s_Z^2 \right)^2}$ &
$\left( \frac{2}{8 \bar s^2 - 3}
+ \frac{6 - 16 \bar s^2}{9 - 24 \bar s^2 + 32 \bar s^4} \right)
\alpha \bar F_Z$
\tabularnewline
\hline 
\end{tabular}\caption{$Z^0$-pole observables.
  They may be found in Table 10.5 of~\cite{particledatagroup2022}.
  Notice that all the $\left. O^\mathrm{NPM} \right/ O^\mathrm{SM}$
  depend only on $\bar s $ and $\alpha \bar F_Z$.
  \label{tab:Z-pole}}
\end{table}

\begin{table}[ht]
\center
\begin{tabular}{|c|c|}
\hline 
Observable $O$ & $\left. O^\mathrm{NPM} \right/ O^\mathrm{SM} - 1$
\tabularnewline
\hline 
\hline 
$\Gamma_{Z\to\ell\bar{\ell}} =
\frac{\widehat \alpha\, m_{Z}}{6 \widehat c^{2}\widehat s^{2}}
\left[ 1 + \Pi_{ZZ}^{\prime}\left(m_{Z}^{2}\right)\right]
\left[ \left( \frac{1}{2} - s_{Z}^{2} \right)^{2}
  + \left(s_{Z}^{2}\right)^{2}\right]$ & 
$\frac{4 \bar s^2 - 1}{8 \bar s^4 -4 \bar s^2 + 1}\, \alpha \bar{F}_Z + c_{TV}$
\tabularnewline
\hline 
$\Gamma_{Z\to\text{inv}}= 3\, \Gamma_{Z\to\nu\bar{\nu}}
=
\frac{ \widehat \alpha\, m_{Z}}{8 \widehat c^{2} \widehat s^{2}}
\left[ 1 + \Pi_{ZZ}^{\prime}\left(m_{Z}^{2}\right) \right]$ & $c_{TV}$
\tabularnewline
\hline 
$\Gamma_{Z\to c\bar{c}} =
\frac{\widehat \alpha\,m_{Z}}{2 \widehat c^{2} \widehat s^{2}}
\left[ 1 + \Pi_{ZZ}^{\prime} \left( m_{Z}^{2} \right) \right]
\left[ \left( \frac{1}{2} - \frac{2}{3}\, s_{Z}^{2} \right)^{2}
  + \left( \frac{2}{3}\, s_{Z}^{2} \right)^{2} \right]$ &
$\frac{16 \bar s^2- 6}{32 \bar s^4 - 24 \bar s^2 + 9}\, \alpha \bar{F}_Z
+ c_{TV}$
\tabularnewline
\hline 
$\Gamma_{Z\to b\bar{b}} =
\frac{\widehat \alpha\,m_{Z}}{2 \widehat c^{2} \widehat s^{2}}
\left[ 1 + \Pi_{ZZ}^{\prime}\left(m_{Z}^{2}\right) \right]
\left[ \left( -\frac{1}{2} +\frac{1}{3}\, s_{Z}^{2}\right)^{2}
  +\left(\frac{1}{3}\, s_{Z}^{2}\right)^{2}\right]$ & 
$\frac{4 \bar s^2 - 3}{8 \bar s^4- 12 \bar s^2 + 9}\, \alpha \bar{F}_Z + c_{TV}$
\tabularnewline
\hline 
$\Gamma_{Z\to\text{hadrons}}
= 2\, \Gamma_{Z\to c\bar{c}} + 3\, \Gamma_{Z\to b\bar{b}}$ & 
$\frac{44 \bar s^2 - 21}{88\bar{s}^{4}-84\bar{s}^{2}+45}\, \alpha \bar{F}_Z
+ c_{TV}$
\tabularnewline
\hline 
$\Gamma_{Z\to\text{all}}=
3\, \Gamma_{Z\to\ell\bar{\ell}} + \Gamma_{Z\to\text{inv}} + \Gamma_{Z\to\text{hadrons}}
$ & 
$\frac{80 \bar s^2 - 30}{160 \bar s^4 - 120 \bar s^2 + 63}\,\alpha \bar{F}_Z
+ c_{TV}$\tabularnewline
\hline 
\end{tabular}
\caption{Observables in the decays of the $Z^0$,
  listed in Table 10.6 of~\cite{particledatagroup2022}.
  Notice that all the $\left. O^\mathrm{NPM} \right/ O^\mathrm{SM}$
  depend only on $\bar s$,
  $\alpha \bar F_Z$,
  and $c_{TV} \equiv \alpha T + \alpha V$.
  \label{tab:Z decays}}
\end{table}

\begin{table}[ht]
\center
\begin{tabular}{|c|c|}
\hline 
Observable $O$ & $\left. O^\mathrm{NPM} \right/ O^\mathrm{SM} - 1$
\tabularnewline
\hline 
\hline 
$g_{A}=-\frac{1}{2}\, \rho_{*}$ & $\alpha T$
\tabularnewline
\hline 
\multirow{1}{*}{$g_{V}=2\rho_{*}\left(s_{0}^{2}-\frac{1}{4}\right)$} &
$\frac{\alpha \bar F_0}{4 \bar s^2 -1} + \alpha T$
\tabularnewline
\hline 
$Q \left(_{\ 55}^{133} \mathrm{Cs} \right)
= - 2 \rho_\ast \left[ 55 \left( 2 s_0^2 - \frac{1}{2} \right) + 39 \right]$ &
$\frac{55\, \alpha \bar F_0}{220\, \bar s^2 + 23} + \alpha T$
\tabularnewline
\hline 
$Q\left(_{\ 81}^{205} \mathrm{Tl} \right)
= - 2 \rho_\ast \left[ 81 \left( 2 s_0^2 - \frac{1}{2}\right) + 62 \right]$ &
$\frac{81\, \alpha \bar F_0}{324\, \bar s^2 + 43} + \alpha T$
\tabularnewline
\hline 
$m_W^2 = \widehat c^2 \widehat m_Z^2$ &
$\frac{\bar s^2 - 1}{2 \bar s^2 - 1}\, \alpha T
+ \frac{\alpha S}{2 \left( 2 \bar s^2 - 1 \right)}
+ \frac{\alpha U}{4 s^2 }$
\tabularnewline
\hline 
$\Gamma_{W} = \frac{\widehat \alpha\, m_{W}}{12\widehat s^{2}}
\left[ 1+\Pi_{WW}^{\prime}\left(m_{W}^{2}\right) \right]$ &
$\frac{\bar s^2 - 1}{2 \bar s^2 - 1}\, \alpha T
+ \frac{\alpha S}{2 \left( 2 \bar s^2 - 1 \right)}
+ \frac{\alpha U}{4 s^2 }
+ \alpha W$
\tabularnewline
\hline 
\end{tabular}
\caption{Other observables, taken from Table~10.4
  of~\cite{particledatagroup2022}. \label{tab:other obs} }
\end{table}

\section{Derivation of the oblique corrections in the case SM \vs\ NPM
  \label{app:NPM vs SM}}

In this appendix we make the explicit derivation of Eqs.~\eqref{mvklof},
with the exception of Eq.~\eqref{eq:rhostar replacement}
which was already derived in the main text,
\cf\ Eq.~\eqref{rhostarSM}.

We write the loop corrections as
\bs
\label{nf93l4}
\ba
\left( m_W^2 \right)^\mathrm{M} &=& \left( \widehat m_W^2 \right)^\mathrm{M}
\left( 1 + R^\mathrm{M}_W \right),
\\
m_Z^2 &=&  \left( \widehat m_Z^2 \right)^\mathrm{M}
\left( 1 + R^\mathrm{M}_Z \right),
\\
\alpha &=& \widehat \alpha^\mathrm{M} \left( 1 + R^\mathrm{M}_\alpha \right),
\\
\gfc &=& \gfchat^\mathrm{M}
\left( 1 + R^\mathrm{M}_G \right).
\ea
\es
The superscript `M' explicitly displays that the quantity depends on the model.
Note that,
since we use the measured $m_Z^2$,
$\alpha$,
and $\gfc$ as the input of renormalization,
those quantities do \emph{not} depend on the model.
We restrict ourselves to renormalization at the one-loop level;
then,
only terms \emph{linear} in $R^\mathrm{M}_W$,
$R^\mathrm{M}_Z$,
$R^\mathrm{M}_\alpha$,
and $ R^\mathrm{M}_G$ must be taken into account.
One has,
according to Eqs.~\eqref{eq:all inputs},
\bs
\label{dv0393}
\ba
R^\mathrm{M}_W &=& \frac{\Pi^\mathrm{M}_{WW} \left( m_W^2 \right) }{m_W^2}
- \frac{2}{sc}\, \frac{\Pi^\mathrm{M}_{ZA} \left( 0 \right)}{m_Z^2} , \\
R^\mathrm{M}_Z &=& \frac{\Pi^\mathrm{M}_{ZZ} \left( m_Z^2 \right)}{m_Z^2}
- \frac{2 c}{s}\, \frac{ \Pi^\mathrm{M}_{ZA} \left( 0 \right)}{m_Z^2} , \\
R^\mathrm{M}_\alpha &=& \Pi_{AA}^{\mathrm{M}\prime} \left( 0 \right), \label{alfa} \\
R^\mathrm{M}_G &=& - \frac{\Pi^\mathrm{M}_{WW} \left( 0 \right)}{m_W^2}
+ \frac{2}{sc}\, \frac{\Pi^\mathrm{M}_{ZA} \left( 0 \right)}{m_Z^2}
+ \delta^\mathrm{M}_{Gc},
\ea
\es
where $\delta^\mathrm{M}_{Gc}$ represents the corrections to $\gfc$
that originate from box diagrams
and from the one-particle-irreducible diagrams
that correct the vertices and the external-fermion legs
in muon decay---\ie,
everything except the oblique corrections.

\subsection{Derivation of the corrections
  to \texorpdfstring{$\bar \rho$}{rho bar}
  and \texorpdfstring{$m_W$}{mW} \label{subsec:rho}} 
In both the SM and the NPM one uses the measured $\alpha$,
$\gfc$,
and $m_Z$ as the input of renormalization.
Following Eq.~\eqref{eq:s2c2 nohat},
the Weinberg angle $\bar \theta_W$ is defined by
\bs
\label{74}
\ba
A &\equiv& \frac{\pi \alpha}{\sqrt{2}\, G_{F(\mathrm{charged})}\, m_Z^2}
\label{74a} \\
&=& \bar c^2 \bar s^2,
\label{74b}
\ea
\es
where $\bar c \equiv \cos{\bar \theta_W}$
and $\bar s \equiv \sin{\bar \theta_W}$.
Clearly,
$A$,
$\bar c$,
and $\bar s$ do \emph{not} depend on the model.
Equation~\eqref{74} means that
\bs
\label{bvjgoff}
\ba
\bar c^2 &=& \frac{1 + \sqrt{1 - 4 A}}{2},
\\
\bar  s^2 &=& \frac{1 - \sqrt{1 - 4 A}}{2}.
\ea
\es
It follows from Eqs.~\eqref{bvjgoff} that
\be
\label{84ikfv9v}
\frac{\mathrm{d} \bar c^2}{\mathrm{d} A} = \frac{1}{\bar  s^2 -\bar c^2}.
\ee

We write
\bs
\label{ff777548}
\ba
\bar c^2 &=& \left(\widehat{c}^{\, 2}\right)^\mathrm{M}
\left( 1 + R^\mathrm{M}_{\bar C} \right), \label{xc} \\
\bar s^2 &=& \left(\widehat{s}^{\, 2}\right)^\mathrm{M}
\left( 1 + R^\mathrm{M}_{\bar S} \right), \label{xs} \\
A &=& \widehat A^\mathrm{M} \left( 1 + R^\mathrm{M}_A \right) \label{xa}, \\
\bar \rho^\mathrm{M} &=& \widehat \rho^{\,\mathrm{M}}
\left( 1 + R^\mathrm{M}_{\bar \rho} \right),
\label{jbkpf}
\ea
\es
where $\bar \rho^\mathrm{M}$ was defined in Eq.~\eqref{eq:s2c2 definition}
and depends on the model through its numerator $\left( m_W^2 \right)^\mathrm{M}$.
Clearly,
\be
R^\mathrm{M}_{\bar \rho} = R^\mathrm{M}_W - R^\mathrm{M}_{\bar C} - R^\mathrm{M}_Z.
\label{849dlf}
\ee

We proceed to calculate $R_{\bar C}^\mathrm{M}$ and $R_{\bar S}^\mathrm{M}$.
We have
\bs
\ba
\bar c^2 &=&\left(\widehat{c}^{\, 2}\right)^\mathrm{M}
\left( 1 + R^\mathrm{M}_{\bar C} \right)
\\ &=& \left(\widehat{c}^{\, 2}\right)^\mathrm{M}
+ \frac{\mathrm{d} \bar{c}^{\, 2}}{\mathrm{d} A}\, \mathrm{d} A
\\ &=& \left(\widehat{c}^{\, 2}\right)^\mathrm{M}
+ \frac{1}{\bar{s}^2 - \bar{c}^2}\, \widehat A^\mathrm{M} R^\mathrm{M}_A
\\ &=& \left(\widehat{c}^{\, 2}\right)^\mathrm{M}
+ \frac{\left(\widehat{c}^{\, 2}\right)^\mathrm{M}
  \left(\widehat{s}^{\, 2}\right)^\mathrm{M} R^\mathrm{M}_A}{\bar s^2 - \bar c^2}.
\ea
\es
Therefore,
\be
R^\mathrm{M}_{\bar C} = \frac{\bar s^2}{\bar s^2 - \bar c^2}\, R^\mathrm{M}_A.
\ee
Using the definition of $A$ in Eq.~\eqref{74a} one obtains
\be
R^\mathrm{M}_{\bar C} = \frac{\bar s^2}{\bar s^2 - \bar c^2}
\left( R^\mathrm{M}_\alpha - R^\mathrm{M}_G - R^\mathrm{M}_Z \right).
\label{nf993}
\ee
From Eqs.~\eqref{xc},
\eqref{xs},
and $\bar c^2 + \bar s^2 =
\left(\widehat{c}^{\, 2}\right)^\mathrm{M}
+ \left(\widehat{s}^{\, 2}\right)^\mathrm{M} = 1$ one has
\be
\bar c^2 R^\mathrm{M}_{\bar C} + \bar s^2 R^\mathrm{M}_{\bar S} = 0.
\label{xcxs}
\ee
Thus,
\be
R^\mathrm{M}_{\bar S} = \frac{\bar c^2}{\bar c^2 - \bar s^2}
\left( R^\mathrm{M}_\alpha - R^\mathrm{M}_G - R^\mathrm{M}_Z \right).
\label{nf994}
\ee
Therefore,
inverting Eq.~\eqref{xs},
\bs
\label{758333}
\ba
\left( \widehat s^{\, 2} \right)^\mathrm{M}
&=& \bar s^2 \left( 1 - R^\mathrm{M}_{\bar S} \right)
\\ &=& \bar s^2 \left[ 1 + \frac{\bar c^2}{\bar s^2 - \bar c^2}
  \left( R^\mathrm{M}_\alpha - R^\mathrm{M}_G - R^\mathrm{M}_Z \right)
  \right].
\ea
\es

We now calculate
\bs
\label{eq: Ra-RG-RZ}
\ba
R^\mathrm{M}_\alpha - R^\mathrm{M}_G - R^\mathrm{M}_Z
&=& \Pi_{AA}^{\mathrm{M}\prime} \left( 0 \right)
+ \frac{\Pi^\mathrm{M}_{WW} \left( 0 \right)}{m_W^2}
- \frac{2}{sc}\, \frac{\Pi^\mathrm{M}_{ZA}(0)}{m_Z^2}
- \delta^\mathrm{M}_{Gc}
\no & &
- \frac{\Pi^\mathrm{M}_{ZZ} \left( m_Z^2 \right)}{m_Z^2}
+ \frac{2 c}{s}\, \frac{\Pi^\mathrm{M}_{ZA} \left( 0 \right)}{m_Z^2}
\\ &=&
\frac{\Pi^\mathrm{M}_{WW} \left( 0 \right)}{m_W^2}
- \frac{\Pi^\mathrm{M}_{ZZ} \left( 0 \right)}{m_Z^2}
- \frac{2 s}{c}\, \frac{\Pi^\mathrm{M}_{ZA} \left( 0 \right)}{m_Z^2}
\no & &
+ \Pi_{AA}^{\mathrm{M}\prime} \left( 0 \right)
- \widetilde \Pi^\mathrm{M}_{ZZ} \left( m_Z^2 \right)
- \delta^\mathrm{M}_{Gc}
\\ &=&
\alpha T^\mathrm{M}
+ \Pi_{AA}^{\mathrm{M}\prime} \left( 0 \right)
- \widetilde \Pi^\mathrm{M}_{ZZ} \left( m_Z^2 \right)
- \delta^\mathrm{M}_{Gc} 
\ea
\es
and
\bs
\label{eq: RW-RZ}
\ba
R^\mathrm{M}_W-R^\mathrm{M}_Z &=&
\frac{\Pi^\mathrm{M}_{WW} \left( m_W^2 \right)}{m_W^2}
- \frac{2}{sc}\, \frac{\Pi^\mathrm{M}_{ZA} \left( 0 \right)}{m_Z^2}
- \frac{\Pi^\mathrm{M}_{ZZ} \left( m_Z^2 \right)}{m_Z^2}
+ \frac{2 c}{s}\, \frac{ \Pi^\mathrm{M}_{ZA} \left( 0 \right)}{m_Z^2}
\\ &=&
\frac{\Pi^\mathrm{M}_{WW} \left( 0 \right) }{m_W^2}
- \frac{\Pi^\mathrm{M}_{ZZ} \left( 0 \right)}{m_Z^2}
- \frac{2 s}{c}\, \frac{ \Pi^\mathrm{M}_{ZA} \left( 0 \right)}{m_Z^2}
\no & &
+ \widetilde \Pi^\mathrm{M}_{WW} \left( m_W^2 \right)
- \widetilde \Pi^\mathrm{M}_{ZZ} \left( m_Z^2 \right)
\\ &=&
\alpha T^\mathrm{M}
+ \widetilde \Pi^\mathrm{M}_{WW} \left( m_W^2 \right)
- \widetilde \Pi^\mathrm{M}_{ZZ} \left( m_Z^2 \right).
\ea
\es

We return to Eq.~\eqref{849dlf}.
Employing Eq.~\eqref{nf993},
\bs
\label{kb0007}
\ba
R^\mathrm{M}_{\bar \rho}
&=& R^\mathrm{M}_W - R^\mathrm{M}_Z + \frac{\bar s^2}{\bar c^2 - \bar s^2}
\left( R^\mathrm{M}_\alpha - R^\mathrm{M}_G - R^\mathrm{M}_Z \right)
\\ &=&
\alpha T^\mathrm{M}
+ \widetilde \Pi^\mathrm{M}_{WW} \left( m_W^2 \right)
- \widetilde \Pi^\mathrm{M}_{ZZ} \left( m_Z^2 \right)
\no & &
+ \frac{\bar s^2}{\bar c^2 - \bar s^2} \left[
  \alpha T^\mathrm{M}
  + \Pi_{AA}^{\mathrm{M}\prime} \left( 0 \right)
  - \widetilde \Pi^\mathrm{M}_{ZZ} \left( m_Z^2 \right)
  - \delta^\mathrm{M}_{Gc} \right]
\\ &=&
\frac{\bar c^2}{\bar c^2 - \bar s^2} \left[
  \alpha T^\mathrm{M}
  - \widetilde \Pi^\mathrm{M}_{ZZ} \left( m_Z^2 \right)
  + \frac{\bar s^2}{\bar c^2}\, \Pi_{AA}^{\mathrm{M}\prime}
  - \frac{\bar s^2}{\bar c^2}\, \delta^\mathrm{M}_{Gc}
  + \frac{\bar c^2 - \bar s^2}{\bar c^2}\,
  \widetilde \Pi^\mathrm{M}_{WW} \left( m_W^2 \right)
  \right]
\\ &=&
\frac{\bar c^2}{\bar c^2 - \bar s^2}
\left( \alpha T^\mathrm{M} - \alpha K^\mathrm{M} \right)
- \frac{\bar s^2}{\bar c^2 - \bar s^2}\, \delta^\mathrm{M}_{Gc}.
\ea
\es

From Eq.~\eqref{jbkpf},
\be
\frac{\bar \rho^\mathrm{NPM}}{\bar \rho^\mathrm{SM}}
= \frac{\widehat \rho^{\,\mathrm{NPM}}}{\widehat \rho^{\,\mathrm{SM}}}
\left( 1 + R^\mathrm{NPM}_{\bar \rho} - R^\mathrm{SM}_{\bar \rho} \right).
\ee
Subtracting Eq.~\eqref{kb0007} in the SM from the same equation in the NPM
and using $\widehat \rho^{\,\mathrm{SM}} = \widehat \rho^{\,\mathrm{NPM}} = 1$,
we arrive at Eq.~\eqref{eq:rhobar NPM-SM};
or,
after neglecting the non-oblique contributions,
at Eq.~\eqref{eq:rhobar SM-NPM oblique}.
As explained in the main text,
that equation is equivalent to Eq.~\eqref{eq:mw NPM vs SM} for $m_W$.
This results in the replacement of Eq.~\eqref{eq:cmz replacement }.

\subsection{Derivation of the corrections
  to \texorpdfstring{$s^2_0$ and $s^2_Z$}{s20 and s2Z} }

In order to derive the corrections to the effective angle,
one starts from
\bs
\ba
\left(s^2_0\right)^\mathrm{M} &=&
\left(\widehat s^2\right)^\mathrm{M}
- \bar s \bar c\, \Pi^{\mathrm{M}\prime}_{ZA} \left( 0 \right),
\label{yue9}
\\
\left(s^2_Z\right)^\mathrm{M} &=&
\left(\widehat s^2\right)^\mathrm{M}
-\bar s \bar c\ \widetilde \Pi^\mathrm{M}_{ZA} \left( m_Z^2 \right)
\\ &=&
\left(s^2_0\right)^\mathrm{M}
-\bar s\bar c \left[
  \widetilde \Pi^\mathrm{M}_{ZA} \left( m_Z^2 \right)
  - \Pi^{\mathrm{M}\prime}_{ZA} \left( 0 \right)
  \right].
\label{uifpds}
\ea
\es
We use Eqs.~\eqref{758333} and~\eqref{eq: Ra-RG-RZ} to get
\bs
\label{nfodd}
\ba
\left(s^2_0\right)^\mathrm{M} &=&
\bar s^2 \left[ 1
  - \frac{\bar c}{\bar s}\ \Pi^{\mathrm{M}\prime}_{ZA} \left( 0 \right)
  + \frac{\bar c^2}{\bar s^2 -\bar c^2}
  \left( R^\mathrm{M}_\alpha - R^\mathrm{M}_G - R^\mathrm{M}_Z \right)
  \right] 
\\ &=&
\bar s^2 \Bigg[ 1
- \frac{\bar c}{ \bar s}\ \Pi^{\mathrm{M}\prime}_{ZA} \left( 0 \right)
+ \frac{\bar c^2}{\bar s^2 -\bar c^2}
\Bigg( \alpha T^\mathrm{M}
  + \Pi_{AA}^{\mathrm{M}\prime} \left( 0 \right)
  - \widetilde \Pi^\mathrm{M}_{ZZ} \left( m_Z^2 \right)
  - \delta^\mathrm{M}_{Gc}  \Bigg)
  \Bigg] 
\\ &=&
\bar s^2 \Bigg[ 1 +\frac{\bar c^2}{\bar s^2 -\bar c^2}
\left( \alpha T^\mathrm{M} + \Pi_{AA}^{\mathrm{M}\prime} \left( 0 \right)
  - \frac{\bar s^2 -\bar c^2}{\bar c \bar s}\,
  \Pi^{\mathrm{M}\prime}_{ZA} \left( 0 \right)
  - \widetilde \Pi^\mathrm{M}_{ZZ} \left( m_Z^2 \right)
  - \delta^\mathrm{M}_{Gc}  \right)
  \Bigg] \hspace*{7mm}
\\ &=& \bar{s}^2\left[
1 
+ \frac{\bar c^2}{\bar s^2 - \bar c^2}\, \alpha T^\mathrm{M}
+ \frac{1}{4 \bar s^2 \left( \bar c^2 - \bar s^2 \right)}\, \alpha S^\mathrm{M}
-\delta^\mathrm{M}_{Gc}
\right].
\ea
\es
Subtracting Eq.~\eqref{nfodd} in the SM from the same equation in an NPM
and using Eqs.~\eqref{eq:S in mass} and~\eqref{za1},
one has
\bs
\label{S0SM}
\ba
\frac{\left( s^2_0 \right)^\mathrm{NPM}}{\left( s^2_0 \right)^\mathrm{SM}}
&=& 1 + \frac{\bar c^2}{\bar s^2 - \bar c^2}\, \alpha T
+ \frac{1}{4 \bar s^2 \left( \bar c^2 - \bar s^2 \right)}\, \alpha S
\\ &=& 1 + \frac{\alpha \bar{F}_0}{4 \bar{s}^2},
\ea
\es
where $\bar{F}_0$ is defined in Eq.~\eqref{fff000}.
Subtracting Eq.~\eqref{uifpds} in the SM from the same equation in an NPM,
one obtains
\be
\label{SZSM}
\frac{\left( s^2_Z \right)^\mathrm{NPM}}{\left( s^2_Z \right)^\mathrm{SM}}
= 1 + \frac{\alpha \bar{F}_Z}{4 \bar{s}^2},
\ee
where $\bar{F}_Z$ is defined in Eq.~\eqref{fffzzz}. 
Equations~\eqref{S0SM} and \eqref{SZSM}
explain the replacement rules in Eqs.~\eqref{eq:s0 replacement}
and~\eqref{eq:sz replacement},
respectively. 

\subsection{Derivation of the corrections to the decay widths
\label{sec:decays} }

The partial decay width of a $W^-$ into an electron and an anti-neutrino is
\be
\Gamma^\mathrm{M} \left( W^- \to e^- \bar \nu_e \right) =
\frac{\widehat \alpha^\mathrm{M}\, m_W}{12
  \left( \widehat s^2 \right)^\mathrm{M}}
\left[ 1 + \Pi^{\mathrm{M}\prime}_{WW} \left( m_W^2 \right) \right].
\label{kfodforr}
\ee
In Eq.~\eqref{kfodforr},
$\widehat \alpha^\mathrm{M}$
and $\left(\widehat s^2\right)^\mathrm{M}$ are bare model parameters,
but $m_W$ is \emph{not}:
it is the physical mass of the $W^-$
and arises from the phase-space integration.
Moreover,
$\Pi^{\mathrm{M}\prime}_{WW} \left( m_W^2 \right)$
originates in the field renormalization.
Inserting the expressions for $\widehat \alpha^\mathrm{M}$
and $\left(\widehat s^2\right)^\mathrm{M}$,
one obtains
\bs
\label{iv0f00}
\ba
\frac{\widehat \alpha^\mathrm{M}}{\left(\widehat s^2\right)^\mathrm{M}}
\left[ 1
  + \Pi^{\mathrm{M}\prime}_{WW} \left( m_W^2 \right) \right]
&=&
\frac{\alpha}{\bar s^2}
\left[ 1 + \Pi^{\mathrm{M}\prime}_{WW} \left( m_W^2 \right)
  - R^\mathrm{M}_\alpha + R^\mathrm{M}_{\bar S} \right]
\\ &=&
\frac{\alpha}{\bar s^2}
\left[ 1 + \Pi^{\mathrm{M}\prime}_{WW} \left( m_W^2 \right)
  - R^\mathrm{M}_\alpha + \frac{\bar c^2}{\bar c^2 - \bar s^2}
  \left( R^\mathrm{M}_\alpha - R^\mathrm{M}_G - R^\mathrm{M}_Z \right) \right]
\hspace*{7mm}
\\ &=&
\frac{\alpha}{\bar s^2}
\left[ \vphantom{\frac{\bar c^2}{\bar c^2 - \bar s^2}}
  1 + \alpha W^\mathrm{M} + \widetilde \Pi_{WW}^\mathrm{M} \left( m_W^2 \right)
  + \frac{\bar s^2}{\bar c^2 - \bar s^2}\,
  \Pi_{AA}^{\mathrm{M}\prime} \left( 0 \right)
  \right. \no & & \left.
  + \frac{\bar c^2}{\bar c^2 - \bar s^2}
  \left( \alpha T^\mathrm{M}
  - \widetilde \Pi^\mathrm{M}_{ZZ} \left( m_Z^2 \right)
  - \delta^\mathrm{M}_{Gc}  \right) \right]
\\ &=&
\frac{\alpha}{\bar{s}^2}\left[1
+ \frac{\bar c^2}{\bar c^2 - \bar s^2}\, \alpha T
+ \frac{\alpha S^\mathrm{M}}{2 \left( \bar s^2 - \bar c^2 \right)}
+ \frac{\alpha U^\mathrm{M}}{4 \bar s^2} + \alpha W^\mathrm{M}
-\delta^\mathrm{M}_{Gc}
\right]
.
\ea
\es
Using Eq.~\eqref{za1} we obtain
\be
\frac{\Gamma^\mathrm{NPM} \left( W^- \to e^- \bar
  \nu_e \right)}{\Gamma^\mathrm{SM} \left( W^- \to e^- \bar
  \nu_e \right)} = 1
+ \frac{\bar c^2}{\bar c^2 - \bar s^2}\, \alpha T
+ \frac{\alpha S}{2 \left( \bar s^2 - \bar c^2 \right)}
+ \frac{\alpha U}{4 \bar s^2} + \alpha W.
\label{pfkkf}
\ee
This leads to the replacement rule of Eq.~\eqref{eq:W decay replacement}. 

Analogously,
for the decay widths of the $Z^0$ we need the following:
\bs
\label{mvkdfo0}
\ba
\frac{\widehat \alpha^\mathrm{M}}{\left(\widehat c^2\right)^\mathrm{M}\,
  \left(\widehat s^2\right)^\mathrm{M}}
\left[ 1 + \Pi^{\mathrm{M}\prime}_{ZZ} \left( m_Z^2 \right) \right]
&=&
\frac{\alpha \left( 1 - R^\mathrm{M}_\alpha \right)}{\bar c^2
  \left( 1 - R^\mathrm{M}_{\bar C} \right)
  \bar s^2 \left( 1 - R^\mathrm{M}_{\bar S} \right)}
\left[ 1 + \Pi^{\mathrm{M}\prime}_{ZZ} \left( m_Z^2 \right) \right]
\\ &=&
\frac{\alpha}{\bar c^2 \bar s^2}
\left[ 1 - R^\mathrm{M}_\alpha + R^\mathrm{M}_{\bar C} + R^\mathrm{M}_{\bar S}
  + \Pi^{\mathrm{M}\prime}_{ZZ} \left( m_Z^2 \right) \right] \hspace{7mm}
\\ &=&
\frac{\alpha}{\bar c^2 \bar s^2}
\left[ 1 - R^\mathrm{M}_G - R^\mathrm{M}_Z
  + \Pi^{\mathrm{M}\prime}_{ZZ} \left( m_Z^2 \right) \right]
\\ &=&
\frac{\alpha}{\bar c^2 \bar s^2}
\left[ 1 + \frac{\Pi^\mathrm{M}_{WW} \left( 0 \right)}{m_W^2}
  -\frac{2}{sc}\, \frac{\Pi^\mathrm{M}_{ZA} \left( 0 \right)}{m_Z^2}
  - \delta^\mathrm{M}_{Gc}
\right. \nonumber \\ && \left. 
- \frac{\Pi^\mathrm{M}_{ZZ} \left( m_Z^2 \right)}{m_Z^2}
+ \frac{2 c}{s}\, \frac{ \Pi^\mathrm{M}_{ZA} \left( 0 \right)}{m_Z^2}
+ \Pi^{\mathrm{M}\prime}_{ZZ} \left( m_Z^2 \right)
\vphantom{\frac{\Pi^\mathrm{M}_{WW} \left( 0 \right)}{m_W^2}}  \right]
 \\ &=&
\frac{\alpha}{\bar c^2 \bar s^2}
\left[ 1
  + \frac{\Pi^\mathrm{M}_{WW} \left( 0 \right)}{m_W^2}
  - \frac{\Pi^\mathrm{M}_{ZZ} \left( 0 \right)}{m_Z^2}
  - \frac{2s}{c}\, \frac{ \Pi^\mathrm{M}_{ZA} \left( 0 \right)}{m_Z^2}
  - \delta^\mathrm{M}_{Gc}
 \right. \no & & \left.
  - \widetilde \Pi^\mathrm{M}_{ZZ} \left( m_Z^2 \right)
  + \Pi^{\mathrm{M}\prime}_{ZZ} \left( m_Z^2 \right)
  \vphantom{\frac{\bar c^2}{\bar s^2 - \bar c^2}} \right]
\\ &=&
\frac{\alpha}{\bar{c}^2\bar{s}^2}\left[
1
+\alpha T^\mathrm{M}
+\alpha V^\mathrm{M}
-\delta^\mathrm{M}_{Gc}
\right].
\ea
\es
This leads to the replacement rule in Eq.~\eqref{eq:Z decay replacement}.

\section{Derivation of the oblique corrections in the case BM \vs\ BBM
 \label{app:BBM vs BM}  }

We consider a Base Model (BM) and a model Beyond the Base Model (BBM),
\ie\ the BM plus extra fermions and/or extra scalars.
We assume that neither the BM nor the BBM have $\rho$ fixed
at the Lagrangian level.
We choose the input observables of renormalization to be $\alpha$,
$\gfc$,
$m_Z$,
and $m_W$.
The Weinberg angle is defined through Eq.~\eqref{eq:s2 nohat}.
Then,
\be
\label{vvofpd}
\rho
= \frac{m_W^2}{c^2 m_Z^2}
= \frac{m_W^2}{\left( 1 - s^2 \right) m_Z^2}
= \frac{\sqrt{2}\, \gfc\, m_W^4}{\left[ \sqrt{2}\, \gfc\, m_W^2
  - \pi \alpha \right] m_Z^2}
\ee
is fully given in terms of the input observables and,
therefore,
it is the same for both models.

\subsection{Derivation of the corrections to \texorpdfstring{$\rho$}{rho}}

We relate $\rho$,
which is fixed by the input observables through Eq.~\eqref{vvofpd},
to the bare $\widehat \rho$.
We utilize once again Eqs.~\eqref{nf93l4} and~\eqref{dv0393}.
From
\bs
\ba
s^2 &=& \frac{\pi \alpha}{\sqrt{2}\, \gfc\, m_W^2},
\\
\left( \widehat s^2 \right)^\mathrm{M} &=&
\frac{\pi \widehat \alpha^\mathrm{M}}{\sqrt{2}\,
  \widehat G^\mathrm{M}_{F(\mathrm{charged})}
  \left( \widehat m^2_W \right)^\mathrm{M}}
\ea
\es
it follows that
\bs
\label{mgjdfiot}
\ba
s^2 &=& \frac{\pi\, \widehat \alpha^\mathrm{M}
  \left( 1 + R^\mathrm{M}_\alpha \right)}{\sqrt{2}\,
  \widehat G^\mathrm{M}_{F(\mathrm{charged})} \left( 1 + R^\mathrm{M}_G \right)
  \left(\widehat m^2_W \right)^\mathrm{M}
  \left( 1 + R^\mathrm{M}_W \right)}
\\ &=&
\left(\widehat s^2 \right)^\mathrm{M}
\left( 1 + R^\mathrm{M}_\alpha - R^\mathrm{M}_G - R^\mathrm{M}_W \right) 
\\ &=& \left(\widehat s^2 \right)^\mathrm{M} \left[ 1
  + \Pi^{\mathrm{M}\prime}_{AA} \left( 0 \right)
  - \widetilde \Pi^\mathrm{M}_{WW} \left( m_W^2 \right)
  - \delta^\mathrm{M}_{Gc}
  \right ]. 
\ea
\es
We write,
just as in Eqs.~\eqref{ff777548},
\bs
\ba
c^2 &=& \left(\widehat{c}^2 \right)^\mathrm{M}
\left( 1 + R^\mathrm{M}_C \right),
\label{RC} \\
s^2 &=& \left(\widehat{s}^2 \right)^\mathrm{M}
\left( 1 + R^\mathrm{M}_S \right).
\label{RS}
\ea
\es
The quantities $R^\mathrm{M}_C$ and $R^\mathrm{M}_S$ obey,
just as in Eq.~\eqref{xcxs},
\be
c^2 R^\mathrm{M}_C + s^2 R^\mathrm{M}_S = 0.
\label{mg90rr}
\ee
Comparing Eqs.~\eqref{mgjdfiot} and~\eqref{RS} we see that
\be
R^\mathrm{M}_S =
 \Pi^{\mathrm{M}\prime}_{AA} \left( 0 \right)
- \widetilde \Pi^\mathrm{M}_{WW} \left( m_W^2 \right)
- \delta^\mathrm{M}_{Gc}.
\label{jgif000}
\ee
Therefore,
from Eq.~\eqref{mg90rr},
\be
R^\mathrm{M}_C = - \frac{s^2}{c^2} \left[
   \Pi^{\mathrm{M}\prime}_{AA} \left( 0 \right)
  - \widetilde \Pi^\mathrm{M}_{WW} \left( m_W^2 \right)
  - \delta^\mathrm{M}_{Gc} \right].
\ee
Then,
\bs
\label{76788549}
\ba
\rho &=& \frac{m_W^2}{c^2 m_Z^2}
\\ &=& \frac{\left(\widehat m^2_W\right)^\mathrm{M}
  \left( 1 + R^\mathrm{M}_W \right)}{\left(\widehat c^2 \right)^\mathrm{M}
  \left( 1 + R^\mathrm{M}_C \right) \left(\widehat m^2_Z\right)^\mathrm{M}
  \left( 1 + R^\mathrm{M}_Z \right)}
\\ &=& \widehat \rho^\mathrm{M} \left( 1 + R^\mathrm{M}_W- R^\mathrm{M}_Z
- R^\mathrm{M}_C \right)
\\ &=& \widehat \rho^\mathrm{M} \Bigg[ 1
+ \alpha T^\mathrm{M}
+ \widetilde \Pi^\mathrm{M}_{WW} \left( m_W^2 \right)
- \widetilde \Pi^\mathrm{M}_{ZZ} \left( m_Z^2 \right)
 \no & &
+ \frac{s^2}{c^2} \left(
   \Pi^{\mathrm{M}\prime}_{AA} \left( 0 \right)
  - \widetilde \Pi^\mathrm{M}_{WW} \left( m_W^2 \right)
  - \delta^\mathrm{M}_{Gc} \right)
\Bigg]
\\ &=& \widehat \rho^\mathrm{M} \Bigg[ 1
+ \alpha T^\mathrm{M}
+ \frac{c^2-s^2}{c^2}\, \widetilde \Pi^\mathrm{M}_{WW} \left( m_W^2 \right)
- \widetilde \Pi^\mathrm{M}_{ZZ} \left( m_Z^2 \right)
+ \frac{s^2}{c^2}\, \Pi^{\mathrm{M}\prime}_{AA}
- \frac{s^2}{c^2}\, \delta^\mathrm{M}_{Gc} \Bigg]
\\ &=&\widehat \rho^\mathrm{M}\left[1+\alpha T^\mathrm{M}-\alpha K^\mathrm{M}-\frac{s^2}{c^2}\delta^\mathrm{M}_{Gc}\right].
\ea
\es
For the final equality we have used the definition of $K^\mathrm{M}$ in Eq.~\eqref{eq:Kustodial par} to arrive at Eq.~\eqref{eq:rho corrections}.

\subsection{Derivation of the corrections
  to \texorpdfstring{$s^2_0$ and $s^2_Z$}{s20 and s2Z}}

We make $\bar s \to s$ and $\bar c \to c$ in Eq.~\eqref{yue9} to obtain
\bs
\ba
\left(s^2_0\right)^\mathrm{M}
&=& \left(\widehat s^2 \right)^\mathrm{M}
- s c\, \Pi^{\mathrm{M}\prime}_{ZA} \left( 0 \right)
\\ &=&
s^2 \left[ 1 - R^\mathrm{M}_S
  - \frac{c}{s}\, \Pi^{\mathrm{M}\prime}_{ZA} \left( 0 \right)
  \right]
\\ &=&
s^2 \left[ 1 
- \Pi^{\mathrm{M}\prime}_{AA} \left( 0 \right) 
+ \widetilde \Pi^\mathrm{M}_{WW} \left( m_W^2 \right)
+ \delta^\mathrm{M}_{Gc}   
  - \frac{c}{s}\ \Pi^{\mathrm{M}\prime}_{ZA} \left( 0 \right)
  \right],
\\ &=& 
s^2\left[1+\frac{\alpha}{4s^2}\left(S^\mathrm{M}+U^\mathrm{M}\right)
    +\delta^\mathrm{M}_{Gc}\right],
\ea
\es
where we have employed Eq.~\eqref{jgif000}.
Neglecting
$\delta_{Gc}^\mathrm{BBM} - \delta_{Gc}^\mathrm{BM}$,
\bs
\label{n7474}
\ba
\frac{\left( s^2_0 \right)^\mathrm{BBM}}{\left( s^2_0 \right)^\mathrm{BM}}
&=& 1 + \frac{\alpha}{4 s^2}\left(S + U\right)
\\
&=& 1 +\frac{\alpha F_0}{4s^2}.
\ea
\es
Now returning to Eq.~\eqref{uifpds}
(with $\bar s \to s$ and $\bar c \to c$)
and employing Eq.~\eqref{eq:X in mass},
we get
\bs
\label{3vjfid}
\ba
\frac{\left( s^2_Z \right)^\mathrm{BBM}}{\left( s^2_Z \right)^\mathrm{BM}}
&=& 1 +\frac{\alpha}{4s^2}\left(S + U + 4 X\right)
\\ &=& 1+\frac{\alpha F_Z}{4s^2}.
\ea
\es

\subsection{Decay widths}

Instead of Eq.~\eqref{iv0f00} we now have
\bs
\ba
\frac{\widehat \alpha^\mathrm{M}}{\left(\widehat s^2\right)^\mathrm{M}}
\left[ 1 + \Pi^{\mathrm{M}\prime}_{WW} \left( m_W^2 \right) \right]
&=&
\frac{\alpha}{s^2}
\left[ 1 + \Pi^{\mathrm{M}\prime}_{WW} \left( m_W^2 \right)
  - R^\mathrm{M}_\alpha + R^\mathrm{M}_S \right] 
\\ &=&
\frac{\alpha}{s^2}
\left[ 1 + \Pi^{\mathrm{M}\prime}_{WW} \left( m_W^2 \right)
  - \widetilde \Pi^\mathrm{M}_{WW} \left( m_W^2 \right)
  - \delta^\mathrm{M}_{Gc} \right]
\\ &=&
\frac{\alpha}{s^2}\left[1+\alpha W^\mathrm{M}-\delta^\mathrm{M}_{Gc}\right].
\ea
\es
Therefore,
neglecting $\delta_{Gc}^\mathrm{BBM} - \delta_{Gc}^\mathrm{BM}$,
\be
\frac{\Gamma^\mathrm{BBM}
  \left( W^- \to e^- \bar \nu_e \right)}{\Gamma^\mathrm{BM}
    \left( W^- \to e^- \bar \nu_e \right)} = 1 + \alpha W.
\label{c49003}
\ee
Similarly,
instead of Eq.~\eqref{mvkdfo0} we now have
\bs
\ba
\frac{\widehat \alpha^\mathrm{M}}{\left( \widehat c^2 \right)^\mathrm{M}\,
  \left( \widehat s^2 \right)^\mathrm{M}}
\left[ 1 + \Pi^{\mathrm{M}\prime}_{ZZ} \left( m_Z^2 \right) \right]
&=& \frac{\alpha}{c^2 s^2}
\left[ 1 - R^\mathrm{M}_\alpha + R^\mathrm{M}_C + R^\mathrm{M}_S + \Pi^{\mathrm{M}\prime}_{ZZ} \left( m_Z^2 \right) \right] 
\\ &=&
\frac{\alpha}{c^2 s^2}
\left[ 1 - R^\mathrm{M}_\alpha
  + \left( 1 - \frac{s^2}{c^2} \right) R^\mathrm{M}_S
  + \Pi^{\mathrm{M}\prime}_{ZZ} \left( m_Z^2 \right) \right]
\\ &=&
\frac{\alpha}{c^2 s^2}
\Bigg[ 1
- \frac{s^2}{c^2}\, \Pi_{AA}^{\mathrm{M}\prime} \left( 0 \right) 
\nonumber \\ & &
+ \frac{s^2 - c^2}{c^2}
\left( \widetilde \Pi^\mathrm{M}_{WW} \left( m_W^2 \right)
  + \delta^\mathrm{M}_{Gc} \right)
+ \Pi^{\mathrm{M}\prime}_{ZZ} \left( m_Z^2 \right)
\Bigg]
\hspace*{7mm} \\  \nonumber &=&
\frac{\alpha}{c^2 s^2} \Bigg[ 1
+ \frac{s^2 - c^2}{c^2}\, \delta^\mathrm{M}_{Gc} 
- \frac{s^2}{c^2}\ \Pi_{AA}^{\mathrm{M}\prime} \left( 0 \right)
\hspace*{7mm} \\ & &
+ \frac{s^2 - c^2}{c^2}\ \widetilde \Pi^\mathrm{M}_{WW} \left( m_W^2 \right)
+ \widetilde \Pi^\mathrm{M}_{ZZ} \left( m_Z^2 \right)
\nonumber \\ & &
+ \left( \Pi^{\mathrm{M}\prime}_{ZZ} \left( m_Z^2 \right)
  - \widetilde \Pi^\mathrm{M}_{ZZ} \left( m_Z^2 \right) \right)
\Bigg]
\\ &=&
\frac{\alpha}{c^2s^2}\left[
1
+\frac{\alpha S^\mathrm{M}}{2c^2}
+\frac{s^2-c^2}{4s^2c^2}\alpha U^\mathrm{M}
+\alpha V^\mathrm{M}
+ \frac{s^2 - c^2}{c^2}\, \delta^\mathrm{M}_{Gc}
\right]. \hspace*{7mm}
\ea
\es
%
We conclude that
\be
\label{w1}
\frac{\Gamma^\mathrm{BBM} \left( Z \to \nu \bar \nu \right)}{\Gamma^\mathrm{BM}
  \left( Z \to \nu \bar \nu \right)} =
1 + \frac{\alpha S}{2 c^2}
+ \frac{s^2 - c^2}{4 s^2 c^2}\, \alpha U + \alpha V.
\ee

\subsection{Summary}

To summarize,
in the confrontation of a BBM with its corresponding BM,
one must use again Tables~\ref{tab:Z-pole},
\ref{tab:Z decays},
and~\ref{tab:other obs},
after making the simple replacements in Eqs.~\eqref{eq:T subs}
and~\eqref{eq: trig sub}.
In particular,
we make
\bs
\label{eq:Fz and F0 no bar}
\ba
\bar F_0 \to F_0 &\equiv&
\frac{S}{c^2 - s^2} + \frac{4 s^2 c^2}{s^2 - c^2}\, K,
\\
\bar F_Z \to F_Z &\equiv& F_0 + 4 X.
\ea
\es
Alternatively,
remembering the definition of $K$ in Eq.~\eqref{eq:T subs},
one obtains
\be
F_0 = S + U.
\ee

\section{Various definitions of the oblique parameters}
\label{app: barbieri}

Originally,
the OPs were defined by assuming NP to lie at a very high energy scale.
A Taylor expansion of the self-energies around $p^2 =0$ was employed,
\viz
\be
\label{84930o}
\Pi_{VV^\prime} \left( p^2 \right)
\approx \Pi_{VV^\prime} \left( 0 \right)
+ p^2\, \Pi^\prime_{VV^\prime} \left( 0 \right)
+ \frac{p^4}{2}\, \Pi^{\prime \prime }_{VV^\prime} \left( 0 \right)+\dots
\ee
Then,
only the $\Pi_{V V^\prime} \left( 0 \right)$
and $\Pi_{V V^\prime}^\prime \left( 0 \right)$ were used for defining $S$,
$T$,
and $U$ in~\cite{peskin1992}. 
%
In order to account for light new physics, Ref.~\cite{maksymyk1994}
did not use the Taylor expansion,
but rather had recourse to modified definitions of $S$ and $U$
and to the inclusion of the three additional oblique parameters $X$,
$V$,
and $W$.
In the limit of heavy NP,
those additional parameters reduce to $\Pi^{\prime \prime }_{VV^\prime} \left( 0 \right)$,
while $S$, $T$ and $U$ also include $\Pi^{\prime \prime }_{VV^\prime} \left( 0 \right)$ in their definitions, as seen in Eqs.~\eqref{v2}--\eqref{x2} below.
In this limit one may relate the OPs of~\cite{maksymyk1994} to the
OPs of \cite{barbieri2004} and thus to the higher-dimensional
coefficients of EFT. 

Utilizing the approximation~\eqref{84930o},
one obtains for the functions defined in Eqs.~\eqref{eq:NP}:
\bs
\label{ug984993}
\ba
\widetilde \Pi_{VV^\prime} \left( p^2 \right)
&\approx&
\Pi^\prime_{VV^\prime} \left( 0 \right)
+ \frac{p^2}{2}\, \Pi^{\prime \prime }_{VV^\prime} \left( 0 \right),
\\
\Pi^\prime_{VV^\prime} \left( p^2 \right) &\approx&
\Pi^\prime_{VV^\prime} \left( 0 \right)
+ p^2\, \Pi^{\prime \prime}_{VV^\prime} \left( 0 \right).
\ea
\es

Plugging Eqs.~\eqref{ug984993} into the definitions of the OPs
in Eqs.~\eqref{eq:STU definitions} gives
\bs
\ba
S & \approx & 
\frac{4 s^2 c^2}{\alpha}
\left[ \Pi^\prime_{ZZ} \left( 0 \right)
  + \frac{m_Z^2}{2}\, \Pi^{\prime\prime}_{ZZ} \left( 0 \right)
  + \frac{s^2 - c^2}{sc}\, \Pi_{ZA}^{\prime } \left( 0 \right)
  - \Pi_{AA}^{\prime} \left( 0 \right)
  \right],
\\
T &=& \frac{1}{\alpha} \left[ \frac{\Pi_{WW}\left( 0 \right)}{m_W^2}
  - \frac{\Pi_{ZZ}\left( 0 \right)}{m_Z^2}
  - \frac{2 s}{c}\, \frac{\Pi_{ZA} \left( 0 \right)}{m_Z^2}
  \right],
\\
U &\approx & 
\frac{4 s^2}{\alpha} \left[
  \Pi^\prime_{WW} \left( 0 \right)
  + \frac{m_W^2}{2}\, \Pi^{\prime \prime}_{WW} \left( 0 \right)
  - c^2\, \Pi^\prime_{ZZ} \left( 0 \right)  \right. \\ 
  && \left.
  - \frac{c^2 m_Z^2}{2}\, \Pi^{\prime\prime}_{ZZ} \left( 0 \right)
  - 2 s c\, \Pi_{ZA}^{\prime} \left( 0 \right)
  - s^2\, \Pi_{AA}^{\prime} \left( 0 \right) \right],
\\
V &\approx & \frac{m_Z^2}{2 \alpha}\, \Pi^{\prime \prime }_{ZZ}\left( 0 \right),
\label{v2} \\
W &\approx& \frac{m_W^2}{2 \alpha}\, \Pi^{\prime \prime }_{WW}\left( 0 \right) ,
\label{w2} \\
X&\approx&
- \frac{s c m_Z^2}{2 \alpha}\, \Pi^{\prime \prime }_{ZA}\left( 0 \right),
\label{x2} \\
Y&  \approx & - \frac{m_Z^2}{2\alpha}\,  \Pi^{\prime\prime}_{AA} \left( 0 \right).
\label{y2}
\ea
\es
Note that now both $S$ and $U$ include second-order derivatives,
in contrast to e.g.~\cite{barbieri2004}.

In order to relate the different definitions in~\cite{peskin1992},
\cite{degrassi1993},
and~\cite{barbieri2004}
it is useful to transform the fields into the gauge eigenstate basis:
\bs
\ba
\Pi_{BB} &=& s^2\, \Pi_{ZZ}  - 2 c s\, \Pi_{ZA} + c^2\, \Pi_{AA}, \\
\Pi_{3 B} &=& - c s\, \Pi_{ZZ} + \left( c^2 - s^2 \right) \Pi_{ZA}
+ c s\, \Pi_{AA}, \\
\Pi_{3 3} &=& c^2\, \Pi_{ZZ} + 2 c s\, \Pi_{ZA} + s^2\, \Pi_{AA},
\ea
\es
%
where all the functions are understood to be evaluated
at the same value of their argument $p^2$.
This leads to
\bs
\label{uuuu34}
\ba
\alpha \left( \frac{S}{4 s c} - s c\,  V \right) &=&
- \Pi^\prime_{3 B} \left( 0 \right),
\\
\alpha m_W^2 T &=&  \Pi_{1 1} \left( 0 \right)
- \rho\,  \Pi_{3 3} \left( 0 \right),
\label{gjie003} \\
\alpha \left( \frac{U}{4 s^2} + c^2 V - W \right) &=&
\Pi^\prime_{1 1} \left( 0 \right) - \Pi^\prime_{3 3} \left( 0 \right),
\\
\alpha W &=& \frac{m_W^2}{2}\, \Pi^{\prime \prime}_{1 1} \left( 0 \right),
\\
\alpha \left( s^2 V + 2 X - c^2 Y \right) &=&
\frac{m_Z^2}{2}\, \Pi^{\prime \prime}_{BB} \left( 0 \right),
\\
\alpha \left( c^2 V - 2 X - s^2 Y \right) &=&
\frac{m_Z^2}{2}\, \Pi^{\prime \prime}_{3 3} \left( 0 \right),
\\
- \frac{\alpha}{c s} \left[ c^2 s^2 \left( V + Y \right)
  + \left( c^2 - s^2 \right) X \right] &=&
\frac{m_Z^2}{2}\, \Pi^{\prime \prime}_{3 B}\left( 0 \right),
\ea
\es
where we have used
\be
\Pi_{11} \left( p^2 \right) = \Pi_{22} \left( p^2 \right)
= \Pi_{WW} \left( p^2 \right),
\ee
and,
in Eq.~\eqref{gjie003},
$\Pi_{AA} \left( 0 \right) = 0$.

We now write the OPs as they were defined in~\cite{barbieri2004},
but firstly we translate
them into a canonically normalized form,
as is also done in~\cite{cacciapaglia2006},
since we use the canonical normalization of fields.
We use the blackboard-bold font to distinguish
our oblique parameters, \ie\ those of~\cite{maksymyk1994},
from the ones of~\cite{barbieri2004}
\bs
\label{uuuu35}
\ba
\frac{\alpha}{4 s c}\ \mathbb S &=& - \Pi^\prime_{3 B} \left( 0 \right),
\\
\alpha m_W^2\, \mathbb T &=&
\Pi_{1 1} \left( 0 \right) -\Pi_{3 3} \left( 0 \right),
\\
\frac{\alpha}{4s^2}\  \mathbb U &=&
\Pi^\prime_{1 1} \left( 0 \right) -\Pi^\prime_{3 3} \left( 0 \right),
\\
\frac{2}{m_W^2}\ \mathbb V &=&
\Pi^{\prime\prime}_{1 1} \left( 0 \right)
- \Pi^{\prime\prime}_{3 3} \left( 0 \right),
\\
 \frac{2}{m_W^2}\ \mathbb X &=& -\Pi^{\prime\prime}_{3 B} \left( 0 \right), \\
\frac{2}{m_W^2}\ \mathbb W &=& - \Pi^{\prime\prime}_{3 3} \left( 0 \right), \\
\frac{2}{m_W^2}\ \mathbb Y &=& - \Pi^{\prime\prime}_{B B} \left( 0 \right).
\ea
\es
Note that,
because~\cite{barbieri2004} absorbed some couplings
and $i$ factors into the fields,
some expressions are different from~\cite{barbieri2004}. 
Assuming $\rho=1$,
one may compare Eqs.~\eqref{uuuu34} and~\eqref{uuuu35} to obtain
\bs
\label{eq:relations between parameters}
\ba
\mathbb S &=& S - 4  s^2 c^2  V,
\\
\mathbb T &=& T,
\\
\mathbb U &=& U + 4 s^2 c^2 V - 4 s^2 W,
\\
\mathbb Y &=& \alpha c^2 \left( - s^2 V - 2 X + c^2 Y \right),
\\
\mathbb W &=& \alpha  c^2 \left( - c^2 V + 2 X + s^2 Y \right), 
\\
\mathbb X &=& \alpha\, \frac{c}{s}
\left[ c^2 s^2 \left( V + Y \right) + \left( c^2 - s^2 \right) X \right],
\\
\mathbb V &=& \mathbb W + \alpha W.
\ea
\es
So,
$Y$ and the six parameters of~\cite{maksymyk1994}
are related to the seven parameters of~\cite{barbieri2004}
in the limit of $\Lambda \gg m_Z$.
In this limit,
one may transform the results of the fit of $\mathbb X$,
$\mathbb W$,
and $\mathbb Y$ in~\cite{barbieri2004} to $V$,
$X$,
and $Y$. 
It is interesting to note that the assumption $Y\approx0$
of~\cite{maksymyk1994} translates into a relation among $\mathbb X$,
$\mathbb W$,
and $\mathbb Y$ which may easily be derived by inverting the transformation
of Eqs.~\eqref{eq:relations between parameters}. 

An insight of~\cite{barbieri2004} is that $\mathbb S$,
$\mathbb T$,
$\mathbb Y$,
and $\mathbb W$ are dominant,
which means that one may assume
$\mathbb U \approx \mathbb V \approx \mathbb X\approx 0$,
producing three restrictions on the $STUVXWY$ parameter set.

Using the translation between the parameters
in Eqs.~\eqref{eq:relations between parameters}
one may also relate the OPs to the Wilson coefficients in SMEFT,
which can be found \textit{e.g.}\ in~\cite{barbieri2004, wells2016}. 
However,
as we stress again,
this only holds for \emph{heavy} NP,
which is also the assumption of SMEFT. 
Thus,
the OPs are more general,
in the sense that they are valid also in parameter regions
where the EFT becomes less accurate.

\section{Ultraviolet finiteness of the oblique parameters}
\label{app:divergences}

The model parameters defined in Eqs.~\eqref{eq:STU definitions},
\ie\ the OPs without the subtraction between two models,
are UV-divergent in a general gauge.
In the comparison between the SM and a NPM,
$SU(2)\times U(1)$ gauge symmetry and the condition $\widehat \rho = 1$
ensure that the subtracted OPs
are finite and gauge-invariant.
In the case of BBM \vs\ BM,
we have used $m_W$ as an additional input observable,
effectively eschewing the oblique parameter $T$. 
While many authors have pointed out that $T$ is divergent
when $\widehat \rho$ is free,
it is interesting to check the structure of its divergence.

We find it insightful to check the UV divergences of the model
(\ie\ \emph{non-subtracted})
parameters in the specific $R_\xi$ gauge that has
\begin{equation}
  \xi_W = \xi_Z = \xi_A \equiv \xi. \label{eq:restricted gauge}
\end{equation}
In this case the theory may be embedded
into the background field formalism
(see for instance~\cite{denner1995},
\cite{denner1995a},
\cite{Denner2020}),
which preserves the global gauge symmetry. 
Ultraviolet divergences may then be more easily tracked,
since UV-divergent operators must preserve the global gauge symmetry. 
As a result,
all the UV divergences are solely absorbed by the redefinitions of the fields,
background fields,
and parameters.
The renormalization of the VEVs is interpreted
as the  renormalization of the Higgs background field~\cite{sperling2013}.

When checking the divergences of the OPs,
it makes no difference whether one uses the functions
$\widetilde {\Pi}_{V V^\prime} \left( p^2 \right)$
or the functions $\Pi^\prime_{V V^\prime} \left( p^2 \right)$ --- defined
in Eqs.~\eqref{eq:NP} --- in the expressions for the OPs,
since these functions collect the same divergent terms,
linear in $p^2$,
of the function ${\Pi}_{V V^\prime} \left( p^2 \right)$.
Having this in mind,
it is clear that the model parameters $V^\mathrm{M}$,
$W^\mathrm{M}$,
and $X^\mathrm{M}$ are automatically finite.
Moreover,
in order to check the UV structure of $S^\mathrm{M}$,
$T^\mathrm{M}$,
and $U^\mathrm{M}$ it is enough to take the limiting case of~\cite{peskin1992},
\ie\ to consider only the first derivatives of the self-energies
in their Taylor expansions.
It is more convenient to use the self-energies
of the gauge eigenstates, 
\ie\ to use the basis of the gauge fields $B_\mu$ and $W_\mu^a$
($a = 1, 2, 3$), by employing eq.~\eqref{eq:Weinberg rotation}, which leads to

\bs
\ba
- \frac{\alpha}{4sc}\, S^\mathrm{M}
&=& \Pi_{3B}^{\prime\, \mathrm{M}} \left( 0 \right),
\label{eq:S in gauge} \\
\alpha m_W^2 T^\mathrm{M} &=& \Pi^\mathrm{M}_{11}\left( 0 \right)
- \rho\, \Pi^\mathrm{M}_{33}\left( 0 \right),
\label{eq:T in gauge} \\
\frac{\alpha}{4 s^2}\, U^\mathrm{M}
&=& \Pi_{11}^{\prime\, \mathrm{M}} \left( 0 \right)
- \Pi_{33}^{\prime\, \mathrm{M}} \left( 0 \right).
\label{eq:U in gauge}
\ea
\es

We consider the model parameters of the model in Section~\ref{sec:BBM model}.
The renormalization transformation in the gauge-eigenstate basis is
\bs
\label{eq:renormalization}  
\ba
\widehat B_\mu &=& \sqrt{1 + \delta_B^\mathrm{M}}\, B_\mu,
\\
\widehat W^a_\mu &=& \sqrt{1 + \delta_W^\mathrm{M}}\, W^a_\mu,
\\
\widehat g &=& \left( 1 + \delta_g^\mathrm{M} \right) g,
\\
\widehat v &=& \left( 1 + \delta_v^\mathrm{M} \right) v, 
\\
\widehat t &=& \left( 1 + \delta_t^\mathrm{M} \right) t,
\\
\widehat u &=& \left( 1 + \delta_u^\mathrm{M} \right) u.
\ea
\es
Notice that there is a single $\delta^\mathrm{M}_W$ for all three $W^a_\mu$.
Furthermore,
there is no mixed $B_\mu$--$W^3_\mu$ renormalization constant,
because of the background-field gauge symmetry. 
Inserting Eqs.~\eqref{eq:renormalization} into the Lagrangian,
one gets the one-loop counterterms for the renormalized self-energies
\bs
\label{eq:self-energies UV}
\ba
\delta \Pi_{11,22}^\mathrm{M} \left( p^2 \right)
&=& - p^2 \delta_W^\mathrm{M}
+ \frac{g^2}{2} \left(\delta \varrho_W^2 \right)^\mathrm{M}
+ \varrho_W^2 \delta_g^\mathrm{M} g^2, \hspace{1cm}
\label{eq:11 in custodial-violating} \\
\delta \Pi_{33}^\mathrm{M} \left( p^2 \right)
&=& - p^2 \delta_W^\mathrm{M}
+ \frac{g^2}{2} \left(\delta \varrho_Z^2 \right)^\mathrm{M}
+ \varrho_Z^2 \delta_g^\mathrm{M} g^2,
\label{eq:33 in custodial-violating}
\ea
\es
where $\delta \varrho_W^2$ and $\delta \varrho_Z^2$
are the variations with respect
to the renormalization transformations in Eq.~\eqref{eq:renormalization}
of the parameters $\varrho_W$ and $\varrho_Z$,
respectively,
defined in Eqs.~\eqref{eq:mw and mz in bbm}. 
Since the divergences in $S^\mathrm{M}$ and $U^\mathrm{M}$
originate in the divergences of the self-energies
that are proportional to $p^2$,
one immediately sees
that $S^\mathrm{M}$ and $U^\mathrm{M}$ are finite
in the gauge defined by Eq.~\eqref{eq:restricted gauge}. 
This holds for all models with $SU(2)\times U(1) $ gauge symmetry,
independently of the value of $\widehat \rho$.

Plugging Eqs.~\eqref{eq:self-energies UV} into Eq.~\eqref{eq:T in gauge},
we gather that the UV divergences of $T^\mathrm{M}$ must be of the form
\bs
\ba
\alpha m_W^2\, \delta T^\mathrm{M} &=&
\frac{g^2}{2} \left[ \left( \delta \varrho_W^2 \right)^\mathrm{M}
  - \frac{\varrho_W^2}{\varrho_Z^2}
  \left( \delta \varrho_Z^2 \right)^\mathrm{M} \right]
\\
&=& \frac{2 g^2}{\varrho_Z^2}
\left[ \left( t^2 - u^2 \right) v^2\, \delta_v^\mathrm{M}
+ \left( 4 t^2 + v^2 \right) u^2\, \delta_u^\mathrm{M}
  - \left( 4 u^2 + v^2 \right) t^2\, \delta_t^\mathrm{M}
\right].
\label{eq:delta T}
\ea
\es
Setting both $u=0$ and $t=0$ in the right-hand side of Eq.~\eqref{eq:delta T}
makes it vanish,
\ie\ one recovers the null result that holds for the SM
(see \eg~\cite{degrassi1993}, \cite{denner1995}). 
But,
$T^\mathrm{M}$ is UV-divergent for either $u \neq 0$ or $t \neq 0$.
Note that $\delta^\mathrm{M}_v$ is different
for M\,$=$\,BBM,
M\,$=$\,BM1,
and M\,$=$\,BM2;
therefore,
in order to obtain a finite subtracted $T$,
one must really have $u=t=0$.
Thus,
the oblique (\ie, subtracted) parameter $T$ is UV-divergent,
with divergences proportional to the VEVs $u$ and $t$
that break the custodial symmetry.
This explains the results previously obtained by many authors,
\eg~\cite{lynn1992,blank1998,czakon2000,forshaw2001,forshaw2003,chen2006a,chankowski2007,chen2008,albergaria2022,rizzo2022}.

\section{Passarino--Veltman functions}
\label{app:PaVe}

Our notation for the Passarino--Veltman functions is
\bs
\ba
\mu^\epsilon \int
\frac{\mathrm{d}^{4 - \epsilon}\, k}{\left( 2 \pi \right)^{4 - \epsilon}}\
\frac{1}{k^2 - I}\ \frac{1}{\left( k + p \right)^2 - J}
&=& \frac{i}{16 \pi^2}\ B_0 \left( P, I, J \right),
\label{jf9995} \\
\mu^\epsilon \int
\frac{\mathrm{d}^{4 - \epsilon}\, k}{\left( 2 \pi \right)^{4 - \epsilon}}\
k^\alpha k^\beta\,
\frac{1}{k^2 - I}\ \frac{1}{\left( k + p \right)^2 - J}
&=& \frac{i}{16 \pi^2} \left[ g^{\alpha \beta}\, B_{00} \left( P, I, J \right)
\right. \nonumber \\ & & \left.
  \qquad + p^\alpha p^\beta\, B_{11} \left( P, I, J \right) \right],
\label{jv9500}
\ea
\es
where $P \equiv p^2$ and in the limit $\epsilon \to 0^+$.
The function $B_{11} \left( P, I, J \right)$ in Eq.~\eqref{jv9500}
is irrelevant for our computations.
The function $B_0 \left( P, I, J \right)$
of Eq.~\eqref{jf9995} only appears in our computations through the combination
\be
F \left( P, I, J \right) \equiv
- B_0 \left( P, I, J \right)
+ \frac{B_{00} \left( P, I, J \right)}{I}.
\ee
Note that $B_0 \left( P, I, J \right)$ and $B_{00} \left( P, I, J \right)$
are symmetric under $I \leftrightarrow J$.

We define
\bs
\ba
B_{00}^\prime \left( P, I, J \right) &\equiv&
\frac{\partial B_{00} \left( P, I, J \right)}{\partial P},
\\
\bar B_{00} \left( P, I, J \right) &\equiv&
\frac{B_{00} \left( P, I, J \right) - B_{00} \left( 0, I, J \right)}{P},
\\
\widetilde B_{00} \left( P, I, J \right) &\equiv&
B_{00}^\prime \left( P, I, J \right) - \bar B_{00} \left( P, I, J \right),
\\
F^\prime \left( P, I, J \right) &\equiv&
\frac{\partial F \left( P, I, J \right)}{\partial P},
\\
\bar F \left( P, I, J \right) &\equiv&
\frac{F \left( P, I, J \right) - F \left( 0, I, J \right)}{P},
\\
\widetilde F \left( P, I, J \right) &\equiv&
F^\prime \left( P, I, J \right) - \bar F \left( P, I, J \right).
\ea
\es

We regularize the Feynman integrals through dimensional regularization.
The divergent quantity
\be
\mathrm{div} \equiv \frac{2}{\epsilon} - \gamma
+ \ln{\left( 4 \pi \right)}
\ee
($\gamma$ is the Euler--Mascheroni constant)
appears in the Feynman integrals in the limit $\epsilon \to 0^+$ as
\bs
\ba
B_{00} \left( P, I, J \right) &=& \left( \frac{I + J}{4}
- \frac{P}{12} \right) \mathrm{div} + \mathrm{finite},
\\
F \left( P, I, J \right) &=& \left( - \frac{3}{4}
+ \frac{J}{4 I} - \frac{P}{12 I} \right) \mathrm{div} + \mathrm{finite}. \hspace{1cm}
\ea
\es
Therefore,
\bs
\label{mvidr00r}
\ba
B_{00}^\prime \left( P, I, J \right) &=& - \frac{\mathrm{div}}{12}
+ \mathrm{finite},
\\
\bar B_{00} \left( P, I, J \right) &=& - \frac{\mathrm{div}}{12}
+ \mathrm{finite},
\\
F^\prime \left( P, I, J \right) &=& - \frac{\mathrm{div}}{12 I}
+ \mathrm{finite},
\\
\bar F \left( P, I, J \right) &=& - \frac{\mathrm{div}}{12 I}
+ \mathrm{finite}.
\ea
\es
Thus,
all four functions $B_{00}^\prime \left( P, I, J \right)$,
$\bar B_{00} \left( P, I, J \right)$,
$I\, F^\prime \left( P, I, J \right)$,
and $I\, \bar F \left( P, I, J \right)$ have identical divergent terms,
and those terms depend neither on $P$ nor on $I$ nor on $J$. 
It also immediately follows that the functions $\tilde{B}_{00}\left(P,I,J\right)$ and $\tilde{F}_{00}\left(P,I,J\right)$ are UV-finite.

\end{appendix}

\pagebreak{}
\bibliographystyle{JHEP}
\bibliography{main.bib}

\providecommand{\href}[2]{#2}\begingroup\raggedright\begin{thebibliography}{10}

\bibitem{kennedy1988}
D.~Kennedy, \emph{Electroweak {{Radiative Corrections}} with an {{Effective
  Lagrangian}}: {{Four-Fermion Processes}}},  Tech. Rep.
  \href{https://www.osti.gov/biblio/1448059}{SLAC-PUB-4039}, {SLAC National
  Accelerator Lab.}, {Menlo Park, CA (United States)} (1988).

\bibitem{veltman1977}
M.~Veltman, \emph{Limit on mass differences in the {{Weinberg}} model},
  \href{https://doi.org/10.1016/0550-3213(77)90342-X}{\emph{Nuclear Physics B}
  {\bfseries 123} (1977) 89}.

\bibitem{peskin1990}
M.E.~Peskin and T.~Takeuchi, \emph{New {{Constraint}} on a {{Strongly
  Interacting Higgs Sector}}},
  \href{https://doi.org/10.1103/PhysRevLett.65.964}{\emph{Physical Review
  Letters} {\bfseries 65} (1990) 964}.

\bibitem{kennedy1990}
D.C.~Kennedy and P.~Langacker, \emph{Precision {{Electroweak Experiments}} and
  {{Heavy Physics}}: {{A Global Analysis}}},
  \href{https://doi.org/10.1103/PhysRevLett.65.2967}{\emph{Physical Review
  Letters} {\bfseries 65} (1990) 2967}.

\bibitem{holdom1990}
B.~Holdom and J.~Terning, \emph{Large corrections to electroweak parameters in
  technicolor theories},
  \href{https://doi.org/10.1016/0370-2693(90)91054-F}{\emph{Physics Letters B}
  {\bfseries 247} (1990) 88}.

\bibitem{golden1991}
M.~Golden and L.~Randall, \emph{Radiative corrections to electroweak parameters
  in technicolor theories},
  \href{https://doi.org/10.1016/0550-3213(91)90614-4}{\emph{Nuclear Physics B}
  {\bfseries 361} (1991) 3}.

\bibitem{peskin1992}
M.E.~Peskin and T.~Takeuchi, \emph{Estimation of {{Oblique Electroweak
  Corrections}}}, \href{https://doi.org/10.1103/PhysRevD.46.381}{\emph{Physical
  Review D} {\bfseries 46} (1992) 381}.

\bibitem{altarelli1991}
G.~Altarelli and R.~Barbieri, \emph{Vacuum {{Polarization Effects}} of {{New
  Physics}} on {{Electroweak Processes}}},
  \href{https://doi.org/10.1016/0370-2693(91)91378-9}{\emph{Physics Letters B}
  {\bfseries 253} (1991) 161}.

\bibitem{altarelli1992}
G.~Altarelli, R.~Barbieri and S.~Jadach, \emph{Toward a {{Model-Independent
  Analysis}} of {{Electroweak Data}}},
  \href{https://doi.org/10.1016/0550-3213(92)90376-M}{\emph{Nuclear Physics B}
  {\bfseries 369} (1992) 3}.

\bibitem{marciano1990}
W.J.~Marciano and J.L.~Rosner, \emph{Atomic parity violation as a probe of new
  physics}, \href{https://doi.org/10.1103/PhysRevLett.65.2963}{\emph{Physical
  Review Letters} {\bfseries 65} (1990) 2963}.

\bibitem{kennedy1991}
D.C.~Kennedy and P.~Langacker, \emph{Erratum: ``{{Precision Electroweak
  Experiments}} and {{Heavy Physics}}: {{A Global Analysis}} [{{Phys}}.
  {{Rev}}. {{Lett}}. 65, 2967 (1990)]},
  \href{https://doi.org/10.1103/PhysRevLett.66.395.2}{\emph{Physical Review
  Letters} {\bfseries 66} (1991) 395}.

\bibitem{maksymyk1994}
I.~Maksymyk, C.P.~Burgess and D.~London, \emph{Beyond {{S}}, {{T}} and {{U}}},
  \href{https://doi.org/10.1103/PhysRevD.50.529}{\emph{Physical Review D}
  {\bfseries 50} (1994) 529}
  [\href{https://arxiv.org/abs/hep-ph/9306267}{{\ttfamily hep-ph/9306267}}].

\bibitem{burgess1994}
C.~Burgess, S.~Godfrey, H.~K{\"o}nig, D.~London and I.~Maksymyk, \emph{A
  {{Global Fit}} to {{Extended Oblique Parameters}}},
  \href{https://doi.org/10.1016/0370-2693(94)91322-6}{\emph{Physics Letters B}
  {\bfseries 326} (1994) 276}.

\bibitem{barbieri2004}
R.~Barbieri, A.~Pomarol, R.~Rattazzi and A.~Strumia, \emph{Electroweak symmetry
  breaking after {{LEP1}} and {{LEP2}}},
  \href{https://doi.org/10.1016/j.nuclphysb.2004.10.014}{\emph{Nuclear Physics
  B} {\bfseries 703} (2004) 127}
  [\href{https://arxiv.org/abs/hep-ph/0405040}{{\ttfamily hep-ph/0405040}}].

\bibitem{wells2016}
J.D.~Wells and Z.~Zhang, \emph{Effective theories of universal theories},
  \href{https://doi.org/10.1007/JHEP01(2016)123}{\emph{Journal of High Energy
  Physics} {\bfseries 2016} (2016) 123}
  [\href{https://arxiv.org/abs/1510.08462}{{\ttfamily 1510.08462}}].

\bibitem{fan2022}
J.~Fan, L.~Li, T.~Liu and K.-F.~Lyu, \emph{W-boson mass, electroweak precision
  tests, and {{SMEFT}}},
  \href{https://doi.org/10.1103/PhysRevD.106.073010}{\emph{Phys. Rev. D}
  {\bfseries 106} (2022) 073010}.

\bibitem{bagnaschi2022}
E.~Bagnaschi, J.~Ellis, M.~Madigan, K.~Mimasu, V.~Sanz and T.~You,
  \emph{{{SMEFT}} analysis of {{mW}}},
  \href{https://doi.org/10.1007/JHEP08(2022)308}{\emph{Journal of High Energy
  Physics} {\bfseries 2022} (2022) 308}.

\bibitem{alonso2014}
R.~Alonso, E.E.~Jenkins, A.V.~Manohar and M.~Trott, \emph{Renormalization group
  evolution of the {{Standard Model}} dimension six operators {{III}}: Gauge
  coupling dependence and phenomenology},
  \href{https://doi.org/10.1007/JHEP04(2014)159}{\emph{Journal of High Energy
  Physics} {\bfseries 2014} (2014) 159}.

\bibitem{henning2017}
B.~Henning, X.~Lu, T.~Melia and H.~Murayama, \emph{2, 84, 30, 993, 560, 15456,
  11962, 261485, . . .: Higher dimension operators in the {{SM EFT}}},
  \href{https://doi.org/10.1007/JHEP08(2017)016}{\emph{Journal of High Energy
  Physics} {\bfseries 2017} (2017) 16}.

\bibitem{banerjee2023}
U.~Banerjee, J.~Chakrabortty, C.~Englert, W.~Naskar, S.U.~Rahaman and
  M.~Spannowsky, \emph{{{EFT}}, {{Decoupling}}, {{Higgs Mixing}} and {{All That
  Jazz}}},  Mar., 2023.

\bibitem{cdfcollaboration+++2022}
{CDF Collaboration{\textdagger}{\textdaggerdbl}}, T.~Aaltonen, S.~Amerio,
  D.~Amidei, A.~Anastassov, A.~Annovi et~al., \emph{High-precision measurement
  of the {{{\emph{W}}}} boson mass with the {{CDF II}} detector},
  \href{https://doi.org/10.1126/science.abk1781}{\emph{Science} {\bfseries 376}
  (2022) 170}.

\bibitem{lynn1992}
B.W.~Lynn and E.~Nardi, \emph{Radiative {{Corrections}} in {{Unconstrained
  SU}}(2) {\texttimes} {{U}}(1) {{Models}} and the {{Top-Mass Problem}}},
  \href{https://doi.org/10.1016/0550-3213(92)90486-U}{\emph{Nuclear Physics B}
  {\bfseries 381} (1992) 467}.

\bibitem{blank1998}
T.~Blank and W.~Hollik, \emph{Precision {{Observables}} in {{SU}}(2) {$\times$}
  {{U}}(1) {{Models}} with an {{Additional Higgs Triplet}}},
  \href{https://doi.org/10.1016/S0550-3213(97)00785-2}{\emph{Nuclear Physics B}
  {\bfseries 514} (1998) 113}
  [\href{https://arxiv.org/abs/hep-ph/9703392}{{\ttfamily hep-ph/9703392}}].

\bibitem{czakon2000}
M.~Czakon, J.~Gluza, F.~Jegerlehner and M.~Zra{\l}ek, \emph{Confronting
  electroweak precision measurements with {{New Physics}} models},
  \href{https://doi.org/10.1007/s100520000278}{\emph{The European Physical
  Journal C} {\bfseries 13} (2000) 275}.

\bibitem{forshaw2001}
J.R.~Forshaw, D.A.~Ross and B.E.~White, \emph{Higgs mass bounds in a {{Triplet
  Model}}}, \href{https://doi.org/10.1088/1126-6708/2001/10/007}{\emph{Journal
  of High Energy Physics} {\bfseries 2001} (2001) 007}
  [\href{https://arxiv.org/abs/hep-ph/0107232}{{\ttfamily hep-ph/0107232}}].

\bibitem{forshaw2003}
J.R.~Forshaw, A.S.~Vera and B.E.~White, \emph{Mass bounds in a model with a
  {{Triplet Higgs}}},
  \href{https://doi.org/10.1088/1126-6708/2003/06/059}{\emph{Journal of High
  Energy Physics} {\bfseries 2003} (2003) 059}
  [\href{https://arxiv.org/abs/hep-ph/0302256}{{\ttfamily hep-ph/0302256}}].

\bibitem{chen2006a}
M.-C.~Chen, S.~Dawson and T.~Krupovnickas, \emph{Constraining new models with
  precision electroweak data},
  \href{https://doi.org/10.1142/S0217751X0603388X}{\emph{International Journal
  of Modern Physics A} {\bfseries 21} (2006) 4045}.

\bibitem{chankowski2007}
P.~Chankowski, S.~Pokorski and J.~Wagner, \emph{({{Non}})decoupling of the
  {{Higgs}} triplet effects},
  \href{https://doi.org/10.1140/epjc/s10052-007-0259-x}{\emph{The European
  Physical Journal C} {\bfseries 50} (2007) 919}.

\bibitem{chen2008}
M.-C.~Chen, S.~Dawson and C.B.~Jackson, \emph{Higgs triplets, decoupling, and
  precision measurements},
  \href{https://doi.org/10.1103/PhysRevD.78.093001}{\emph{Physical Review D}
  {\bfseries 78} (2008) 093001}.

\bibitem{albergaria2022}
F.~Albergaria and L.~Lavoura, \emph{Prescription for finite oblique parameters
  {{S}} and {{U}} in extensions of the {{SM}} with {$m_W\neq
  m_Z\cos\theta_W$}},
  \href{https://doi.org/10.1088/1361-6471/ac7a56}{\emph{Journal of Physics G:
  Nuclear and Particle Physics} {\bfseries 49} (2022) 085005}.

\bibitem{rizzo2022}
T.G.~Rizzo, \emph{Kinetic {{Mixing}}, {{Dark Higgs Triplets}}, {$M_W$} and
  {{All That}}},
  \href{https://doi.org/10.1103/PhysRevD.106.035024}{\emph{Physical Review D}
  {\bfseries 106} (2022) 035024}
  [\href{https://arxiv.org/abs/2206.09814}{{\ttfamily 2206.09814}}].

\bibitem{kennedy1989}
D.~Kennedy and B.~Lynn, \emph{Electroweak radiative corrections with an
  effective lagrangian: {{Four-fermions}} processes},
  \href{https://doi.org/10.1016/0550-3213(89)90483-5}{\emph{Nuclear Physics B}
  {\bfseries 322} (1989) 1}.

\bibitem{cheng2023}
Y.~Cheng, X.-G.~He, F.~Huang, J.~Sun and Z.-P.~Xing, \emph{Electroweak
  precision tests for triplet scalars},
  \href{https://doi.org/10.1016/j.nuclphysb.2023.116118}{\emph{Nuclear Physics
  B} {\bfseries 989} (2023) 116118}
  [\href{https://arxiv.org/abs/2208.06760}{{\ttfamily 2208.06760}}].

\bibitem{song2023}
H.~Song, X.~Wan and J.-H.~Yu, \emph{Custodial symmetry violation in scalar
  extensions of the standard model*},
  \href{https://doi.org/10.1088/1674-1137/ace5a6}{\emph{Chinese Physics C}
  {\bfseries 47} (2023) 103103}.

\bibitem{burdman2008}
G.~Burdman and L.~Da~Rold, \emph{Renormalization of the {{S Parameter}} in
  {{Holographic Models}} of {{Electroweak Symmetry Breaking}}},
  \href{https://doi.org/10.1088/1126-6708/2008/11/025}{\emph{Journal of High
  Energy Physics} {\bfseries 2008} (2008) 025}
  [\href{https://arxiv.org/abs/0809.4009}{{\ttfamily 0809.4009}}].

\bibitem{chun2012}
E.J.~Chun, H.M.~Lee and P.~Sharma, \emph{Vacuum {{Stability}},
  {{Perturbativity}}, {{EWPD}} and {{Higgs-to-diphoton}} rate in {{Type II
  Seesaw Models}}},
  \href{https://doi.org/10.1007/JHEP11(2012)106}{\emph{Journal of High Energy
  Physics} {\bfseries 2012} (2012) 106}
  [\href{https://arxiv.org/abs/1209.1303}{{\ttfamily 1209.1303}}].

\bibitem{mandal2022}
S.~Mandal, O.G.~Miranda, G.S.~Garcia, J.W.F.~Valle and X.-J.~Xu, \emph{Towards
  deconstructing the simplest seesaw mechanism},
  \href{https://doi.org/10.1103/PhysRevD.105.095020}{\emph{Physical Review D}
  {\bfseries 105} (2022) 095020}
  [\href{https://arxiv.org/abs/2203.06362}{{\ttfamily 2203.06362}}].

\bibitem{hagiwara1994}
K.~Hagiwara, D.~Haidt, C.S.~Kim and S.~Matsumoto, \emph{A novel approach to
  confront electroweak data and theory},
  \href{https://doi.org/10.1007/BF01957770}{\emph{Zeitschrift f{\"u}r Physik C
  Particles and Fields} {\bfseries 64} (1994) 559}.

\bibitem{heeck2022}
J.~Heeck, \emph{W-boson mass in the triplet seesaw model},
  \href{https://doi.org/10.1103/PhysRevD.106.015004}{\emph{Physical Review D}
  {\bfseries 106} (2022) 015004}
  [\href{https://arxiv.org/abs/2204.10274}{{\ttfamily 2204.10274}}].

\bibitem{denner1993}
A.~Denner, \emph{{Techniques for the Calculation of Electroweak Radiative
  Corrections at the One-Loop Level and Results forW-physics at LEP 200}},
  \href{https://doi.org/10.1002/prop.2190410402}{\emph{Fortschritte der
  Physik/Progress of Physics} {\bfseries 41} (1993) 307}.

\bibitem{pokorski2000}
S.~Pokorski, \emph{Gauge Field Theories}, Cambridge Monographs on Mathematical
  Physics, {Cambridge University Press}, {Cambridge, U.K. ; New York}, 2nd
  ed~ed. (2000).

\bibitem{bohm2001}
M.~B{\"o}hm, A.~Denner and H.~Joos, \emph{Gauge {{Theories}} of the {{Strong}}
  and {{Electroweak Interaction}}}, {Vieweg+Teubner Verlag}, {Wiesbaden}
  (2001),
  \href{https://doi.org/10.1007/978-3-322-80160-9}{10.1007/978-3-322-80160-9}.

\bibitem{Denner2020}
A.~Denner and S.~Dittmaier, \emph{Electroweak radiative corrections for
  collider physics},
  \href{https://doi.org/10.1016/j.physrep.2020.04.001}{\emph{Physics Reports}
  {\bfseries 864} (2020) 1} [\href{https://arxiv.org/abs/1912.06823}{{\ttfamily
  1912.06823}}].

\bibitem{altarelli1989}
G.~Altarelli, R.H.P.~Kleiss and C.~Verzegnassi, \emph{Z {{Physics}} at
  {{LEP1}}},  Tech. Rep.
  \href{http://cds.cern.ch/record/116932}{CERN-89-08-V-1}, {CERN}, {Geneva,
  Switzerland} (1989), \href{https://doi.org/10.5170/CERN-1989-008-V-1}{DOI}.

\bibitem{Aoki1982}
K.-i.~Aoki, Z.~Hioki, R.~Kawabe, M.~Konuma and T.~Muta, \emph{Electroweak
  {{Theory}}: {{Framework}} of {{On-Shell Renormalization}} and {{Study}} of
  {{Higher-Order Effects}}},
  \href{https://doi.org/10.1143/PTPS.73.1}{\emph{Progress of Theoretical
  Physics Supplement} {\bfseries 73} (1982) 1}.

\bibitem{particledatagroup2022}
{Particle Data Group}, R.L.~Workman, V.D.~Burkert, V.~Crede, E.~Klempt,
  U.~Thoma et~al., \emph{Review of {{Particle Physics}}},
  \href{https://doi.org/10.1093/ptep/ptac097}{\emph{Progress of Theoretical and
  Experimental Physics} {\bfseries 2022} (2022) 083C01}.

\bibitem{sirunyan2019}
A.~Sirunyan, A.~Tumasyan, W.~Adam, F.~Ambrogi, E.~Asilar, T.~Bergauer et~al.,
  \emph{Search for a standard model-like {{Higgs}} boson in the mass range
  between 70 and 110 {{GeV}} in the diphoton final state in proton-proton
  collisions at s = 8 and 13 {{TeV}}},
  \href{https://doi.org/10.1016/j.physletb.2019.03.064}{\emph{Physics Letters
  B} {\bfseries 793} (2019) 320}.

\bibitem{thecmscollaboration2023}
{The CMS collaboration}, A.~Tumasyan, W.~Adam, J.W.~Andrejkovic, T.~Bergauer,
  S.~Chatterjee et~al., \emph{Searches for additional {{Higgs}} bosons and for
  vector leptoquarks in {$\tau\tau$} final states in proton-proton collisions
  at {$ \sqrt{s} = 13~\mathrm{TeV}$}},
  \href{https://doi.org/10.1007/JHEP07(2023)073}{\emph{Journal of High Energy
  Physics} {\bfseries 2023} (2023) 73}.

\bibitem{gascon-shotkinsusan2023}
{Gascon-Shotkin, Susan}, \emph{Searches for additional {{Higgs}} bosons at low
  mass},  Mar., 2023.

\bibitem{thecmscollaboration2023a}
{\scshape CMS} collaboration, \emph{Search for a standard model-like {{Higgs}}
  boson in the mass range between 70 and {$110~\mathrm{GeV}$} in the diphoton
  final state in proton-proton collisions at {$\sqrt{s}=13~\mathrm{TeV}$}},
  Tech. Rep. \href{https://cds.cern.ch/record/2852907}{CMS-PAS-HIG-20-002},
  {CERN}, {Geneva} (2023).

\bibitem{ashanujjaman2023}
S.~Ashanujjaman, S.~Banik, G.~Coloretti, A.~Crivellin, B.~Mellado and
  A.-T.~Mulaudzi, \emph{{$SU(2)_L$} triplet scalar as the origin of the 95
  {{GeV}} excess?},  June, 2023.

\bibitem{lynn1989}
B.W.~Lynn, \emph{{{GAUGE INVARIANCE OF RUNNING COUPLINGS IN}} {$SU(2)_L \times
  U(1)$}},  Tech. Rep. \href{http://cds.cern.ch/record/201416}{SLAC-PUB-5077,
  SU-ITP-86-7}, {SLAC National Accelerator Lab.}, {Stanford, CA} (May, 1989).

\bibitem{degrassi1993}
G.~Degrassi, B.A.~Kniehl and A.~Sirlin, \emph{Gauge-{{Invariant Formulation}}
  of the {$S$}, {$T$}, and {$U$} {{Parameters}}},
  \href{https://doi.org/10.1103/PhysRevD.48.R3963}{\emph{Physical Review D}
  {\bfseries 48} (1993) R3963}.

\bibitem{cacciapaglia2006}
G.~Cacciapaglia, C.~Cs{\'a}ki, G.~Marandella and A.~Strumia, \emph{The
  {{Minimal Set}} of {{Electroweak Precision Parameters}}},
  \href{https://doi.org/10.1103/PhysRevD.74.033011}{\emph{Physical Review D}
  {\bfseries 74} (2006) 033011}
  [\href{https://arxiv.org/abs/hep-ph/0604111}{{\ttfamily hep-ph/0604111}}].

\bibitem{denner1995}
A.~Denner, S.~Dittmaier and G.~Weiglein, \emph{Gauge invariance,
  gauge-parameter independence and properties of {{Green}} functions},  in
  \emph{Ringberg {{Workshop}} on {{Perspectives}} for Electroweak Interactions
  in E+e- Collisions}, ({Tegernesee, Germany}), pp.~281--304, May, 1995,
  \href{http://arxiv.org/abs/hep-ph/9505271}{http://arxiv.org/abs/hep-ph/9505271}
  [\href{https://arxiv.org/abs/hep-ph/9505271}{{\ttfamily hep-ph/9505271}}].

\bibitem{denner1995a}
A.~Denner, G.~Weiglein and S.~Dittmaier, \emph{Application of the
  background-field method to the electroweak standard model},
  \href{https://doi.org/10.1016/0550-3213(95)00037-S}{\emph{Nuclear Physics B}
  {\bfseries 440} (1995) 95}.

\bibitem{sperling2013}
M.~Sperling, D.~St{\"o}ckinger and A.~Voigt, \emph{Renormalization of {{Vacuum
  Expectation Values}} in {{Spontaneously Broken Gauge Theories}}},
  \href{https://doi.org/10.1007/JHEP07(2013)132}{\emph{Journal of High Energy
  Physics} {\bfseries 2013} (2013) 132}
  [\href{https://arxiv.org/abs/1305.1548}{{\ttfamily 1305.1548}}].

\end{thebibliography}\endgroup
\end{document}